\documentclass[twocolumn]{aastex63}

\received{18 September 2020}
\revised{11 December 2020}
\accepted{15 December 2020}
\submitjournal{PSJ}

\shorttitle{Heavy positive ion groups in Titan's ionosphere: Cassini Plasma Spectrometer IBS observations}
\shortauthors{Haythornthwaite et al.}

\begin{document}

\title{Heavy Positive Ion Groups in Titan's Ionosphere from Cassini Plasma Spectrometer IBS Observations}

\correspondingauthor{Richard Haythornthwaite}
\email{richard.haythornthwaite.18@ucl.ac.uk, richardhaythornthwaite@hotmail.co.uk}

\author[0000-0002-5836-2827]{Richard P. Haythornthwaite}
\affiliation{Mullard Space Science Laboratory, Department of Space and Climate Physics, University College London, Dorking, UK}
\affiliation{The Centre for Planetary Sciences at UCL/Birkbeck, London, UK}

\author[0000-0002-6185-3125]{Andrew J. Coates}
\affiliation{Mullard Space Science Laboratory, Department of Space and Climate Physics, University College London, Dorking, UK}
\affiliation{The Centre for Planetary Sciences at UCL/Birkbeck, London, UK}

\author[0000-0002-5859-1136]{Geraint H. Jones}
\affiliation{Mullard Space Science Laboratory, Department of Space and Climate Physics, University College London, Dorking, UK}
\affiliation{The Centre for Planetary Sciences at UCL/Birkbeck, London, UK}

\author[0000-0002-2861-7999]{Anne Wellbrock}
\affiliation{Mullard Space Science Laboratory, Department of Space and Climate Physics, University College London, Dorking, UK}
\affiliation{The Centre for Planetary Sciences at UCL/Birkbeck, London, UK}

\author[0000-0002-1978-1025]{J. Hunter Waite}
\affiliation{Space Science and Engineering Division, Southwest Research Institute, San Antonio, Texas, 78228, USA}

\author[0000-0001-7273-1898]{V\'{e}ronique Vuitton}
\affiliation{Institut de Plan\'{e}tologie et d'Astrophysique de Grenoble, Univ. Grenoble Alpes, CNRS, Grenoble 38000, France}

\author[0000-0002-5360-3660]{Panayotis Lavvas}
\affiliation{Universit\'{e} de Reims Champagne Ardenne, CNRS, GSMA UMR 7331, 51097 Reims, France}

\nocollaboration{7}

\begin{abstract}

Titan's ionosphere contains a plethora of hydrocarbons and nitrile cations and anions as measured by the Ion Neutral Mass Spectrometer and Cassini Plasma Spectrometer (CAPS) onboard the Cassini spacecraft.
Data from the CAPS Ion Beam Spectrometer (IBS) sensor have been examined for five close encounters of Titan during 2009.
The high relative velocity of Cassini with respect to the cold ions in Titan's ionosphere allows CAPS IBS to function as a mass spectrometer.
Positive ion masses between 170 and 310 u/q are examined with ion mass groups identified between 170 and 275 u/q containing between 14 and 21 heavy (carbon/nitrogen/oxygen) atoms.
These groups are the heaviest positive ion groups reported so far from the available \textit{in situ} ion data at Titan.
The ion group peaks are found to be consistent with masses associated with Polycyclic Aromatic Compounds (PAC), including Polycyclic Aromatic Hydrocarbon (PAH) and nitrogen-bearing polycyclic aromatic molecular ions.
The ion group peak identifications are compared with previously proposed neutral PAHs and are found to be at similar masses, supporting a PAH interpretation.
The spacing between the ion group peaks is also investigated, finding a spacing of 12 or 13 u/q indicating the addition of C or CH.
Lastly, the occurrence of several ion groups is seen to vary across the five flybys studied, possibly relating to the varying solar radiation conditions observed across the flybys.
These findings further the understanding between the low mass ions and the high mass negative ions, as well as with aerosol formation in Titan's atmosphere.
\end{abstract}



\section{Introduction} \label{sec:intro}

Titan is the largest moon of Saturn and the only known moon in the solar system with a dense atmosphere.
Since its discovery by Christiaan Huygens it has been observed using both telescopes and spacecraft including the Pioneer 11, Voyager 1 \& 2 spacecraft and most recently Cassini-Huygens. 
Data from the Voyager flybys revealed Titan to possess a thick atmosphere (150 kPa) rich in nitrogen and hydrocarbon compounds \citep{Tyler1981,Lindal1981}.
During Cassini's tour of the Saturnian system between 2004 and 2017 the Cassini spacecraft performed over 100 flybys of Titan, taking in-situ measurements of its ionosphere and upper atmosphere.
These in-situ measurements helped demonstrate the connection between ion-neutral reactions generating large charged molecules in the upper atmosphere and the formation of aerosols \citep{Waite2005,Waite2007,Coates2007}.

Titan's atmosphere is nitrogen-rich (95-98\%) and contains minor species of methane (2-5\%) and hydrogen (0.1\%).
A plethora of trace species exist including hydrocarbons, nitriles and oxygen-bearing molecules such as carbon dioxide and carbon monoxide \citep{Coustenis2007,Waite2005}.
The oxygen-bearing molecules are thought to originate from O\textsuperscript{+} ions and micrometeorites entering Titan's atmosphere \citep{Hartle2006,Horst2012}.

Noticeable features of Titan's atmosphere are the haze layers that give rise to the moon's orange appearance.
The layers are optically thick in most visible wavelengths making remote observations of Titan's surface difficult \citep{Porco2005}.
The organic haze is made up of aerosols and from early experiments using N\textsubscript{2}/CH\textsubscript{4} mixtures, a combination of polycyclic aromatic hydrocarbons (PAHs) and polycyclic aromatic nitrogen heterocycles (PANHs) was proposed as the composition of the aerosols \citep{Sagan1993}.
The detection of benzene from both remote sensing \citep{Coustenis2003} and \textit{in situ} \citep{Waite2007} measurements supported this interpretation, benzene being an aromatic precursor to larger PAHs.
Early Titan flybys identified possible heavy PAH ions such as naphthalene, anthracene derivatives and an anthracene dimer at 130, 170 and 335 u/q respectively \citep{Waite2007} and these are thought to be formed through ion-molecule reactions in the upper atmosphere \citep{Westlake2014}.
Remote sensing infrared measurements have also suggested the presence of PAHs/PANHS up to several hundred amu in the atmosphere \citep{Lopez-Puertas2013}.

Previous ion composition studies by the Ion and Neutral Mass Spectrometer (INMS) and Cassini Plasma Spectrometer (CAPS) instruments revealed ``families'' of ions around particular mass values and a regular spacing of 12 to 14 u/q between mass group peaks characteristic of spectra of complex organic compounds and related to a carbon or nitrogen backbone that dominates the ion chemistry \citep{Waite2005,Cravens2006,Vuitton2007,Crary2009}.
Early Titan flybys by Cassini revealed positive ion masses up to 350 u/q, obtained using measurements from the CAPS Ion Beam Spectrometer \citep{Waite2007}.
Later flybys demonstrated the existence of even larger positive ions up to 1100 u/q \citep{Coates2010-RSC}.
In addition to positive ions, negative ions have been detected in Titan's ionosphere \citep{Coates2007}.
These negative ions extend to higher masses than the positive ions with masses up to 13,800 u/q \citep{Coates2009,Wellbrock2013,Wellbrock2019}.
The negative ions are thought to be composed of carbon chains, with the higher masses having a suspected PAH/PANH composition \citep{Coates2007,Desai_2017}.

The studies by \citet{Crary2009}; \citet{Mandt2012} and \citet{Westlake2014} used IBS data to study various aspects of ion composition at Titan.
These compositional studies covered positive ions (1-200 u/q) in Titan's ionosphere and showed that the likely origin of the ions above 100 u/q were aromatic compounds created through ion-molecule reactions.
\citet{Crary2009} discuss possible ion composition for mass groups up to C13, where C13 indicates the mass group where there are 13 ``heavy'' (carbon, nitrogen or oxygen) atoms present.
The heaviest mass group they reported was the C15 group although the possible ion composition of this material was not discussed. 

There have been several experimental efforts to replicate the organic haze present at Titan along with measurements of the ion-molecule gas-phase chemistry \citep{Dubois2020}.
Some of these experiments have measured products up to 400 amu \citep{Berry2019b}.
Chemical models based on CAPS measurements have proposed molecules up to the C17 group using a mechanistic approach to explain the carbocation/carboanion chemistry \citep{Ali2015}.
The schemes studied included hydrocarbons and nitriles and various reaction paths for the growth of these molecules.

Ions with masses between 170 and 310 u/q are examined in this study with ion mass groups identified up to 275 u/q.
Prominent groups identified contain 15, 16, 18 and 21 heavy atoms.
Several ion groups also show variation in their occurrence between flybys.

\section{Methodology} \label{sec:Methodology}

\begin{figure*}
    \includegraphics[width=1\textwidth]{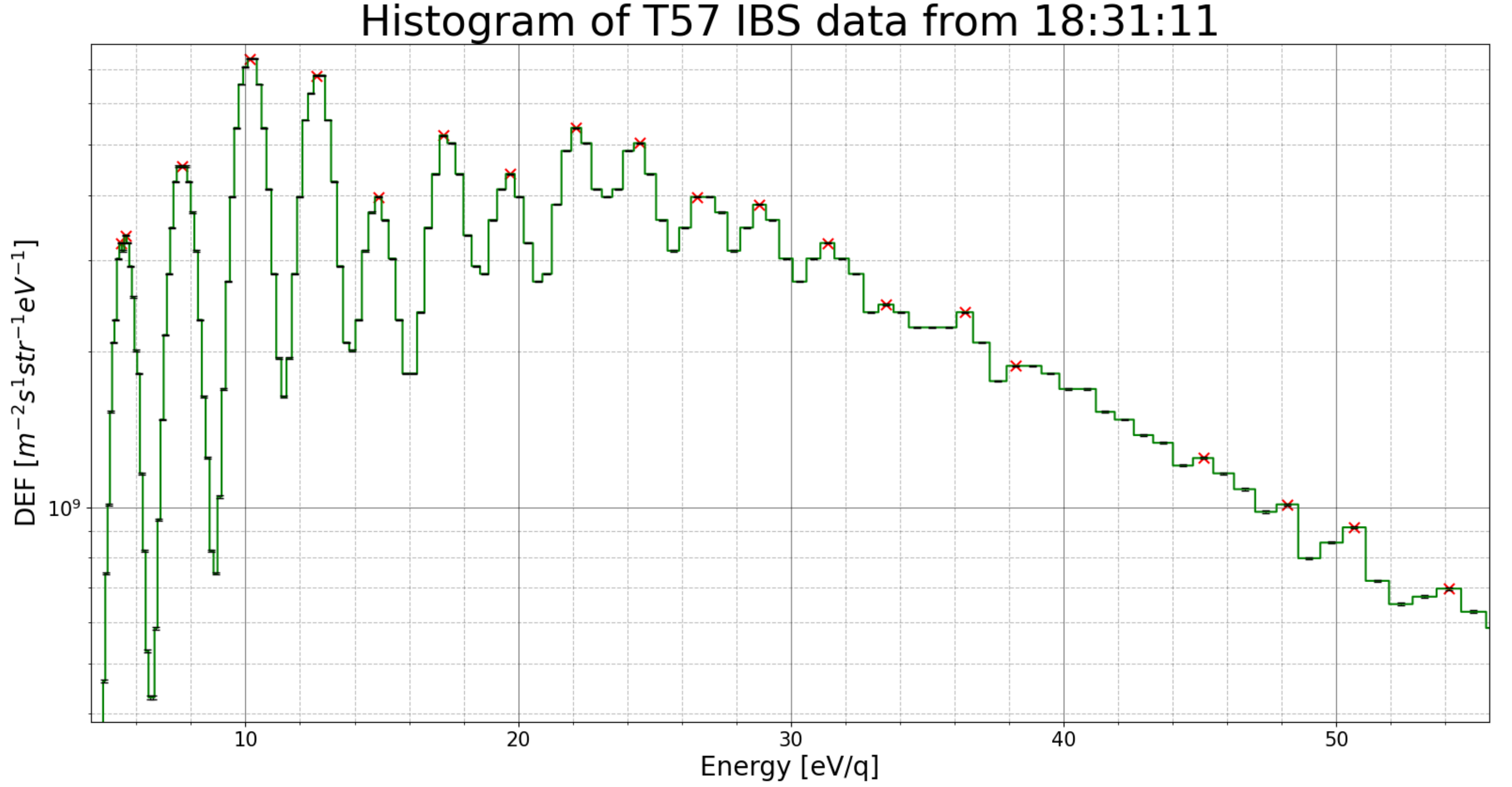}
    \caption{An example of an IBS energy spectra during the T57 flyby. The error bars shown represent the uncertainty due to Poisson counting error. Red x's indicate the peaks as identified by the peak finding algorithm. Due to the logarithmic energy scale, at low energies the ion beams can be seen over a number of bins, while at the high masses the beams can only be seen in a single bin.}
    \label{fig:example_spectra}
\end{figure*}

\subsection{Cassini Plasma Spectrometer} \label{sec:CAPS}
The Cassini spacecraft carried a full suite of scientific instruments including fields, waves, remote sensing, neutral and ion particle instruments.
One of these instruments was the CAPS instrument which studied ions and electrons using three electrostatic analyzers mounted on an actuating platform. 
CAPS was in operation up to 2011 and a short time during 2012.
The three CAPS sensors were the Electron Spectrometer (ELS), Ion Mass Spectrometer (IMS) and the Ion Beam Spectrometer (IBS). 

Only data from CAPS IBS were used for the present study.
CAPS IBS is a curved-electrode electrostatic analyzer and measures ion flux as a function of kinetic energy and direction \citep{Young2004}. 
The ions enter the sensor through three entrance fans (Field of View, $150^{\circ} \times 1.4^{\circ}$) which are tilted at $30^{\circ}$ to each other.
The energy resolution of IBS is 0.014 $\Delta E/E$ and during the Titan encounters, the sensors swept from 207 eV down to 3 eV through 255 adjacent energy bins \citep{Young2004}. 
As CAPS is mounted on an actuating platform, IBS can be rotated to increase its field of view but for the interest of this study we have focused on the Titan flybys where CAPS was at fixed actuation. 
This direction is approximately in the -X direction in the spacecraft frame and is aligned with the view direction of INMS.
By using flybys where CAPS has a fixed actuation in the ram direction more data points can be obtained throughout the flyby.

\subsection{Data Analysis and Ion Winds} \label{sec:Data analysis}
\begin{figure*}
    \includegraphics[width=1\textwidth]{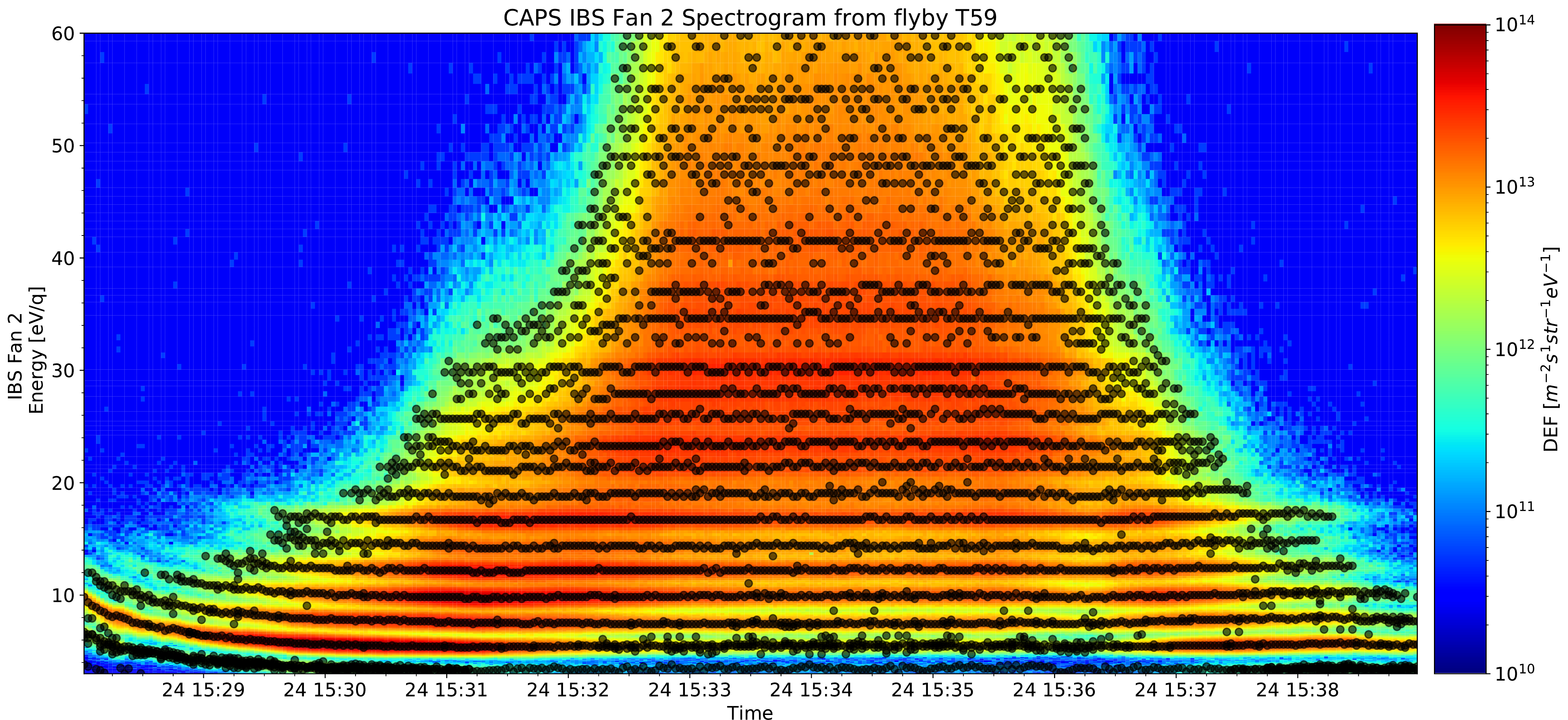}
    \caption{An energy/charge spectrogram from the T59 Titan flyby with intensity shown in differential energy flux. Peaks identified by the algorithm are over plotted with dots. The mass groups identified can clearly be seen by the horizontal lines of dots. Small variations in the energy of the identified peaks are due to effects such as ion velocity, spacecraft potential and slight shifts in spacecraft orientation. The energy shift on the inbound leg of the flyby (before 15:30) is due to the acceleration of ions away from Titan.}
    \label{fig:IBST59PeakAnalysis}
\end{figure*}

The first stage of the analysis was running a peak finding algorithm on each IBS energy sweep over a flyby.
An example of one energy sweep can be seen in Figure \ref{fig:example_spectra}.
A peak was defined as any bin with a count number above the noise level of the instrument as well as above the Poisson counting error, i.e. the square root of the count number, when compared with adjacent energy bins.
This was repeated across a flyby, Figure \ref{fig:IBST59PeakAnalysis} shows the peaks identified during the T59 flyby over plotted on a spectrogram. 

During the Titan flybys, Cassini  had a high velocity (6 km/s)  relative to the low ion thermal (150K) and wind velocities (\textless 230 m/s) in the ionosphere \citep{Crary2009}. 
The combination of these factors means that the ions appear as a highly-directed cold supersonic beam in the spacecraft frame.
The ions therefore appear at kinetic energies associated with the spacecraft velocity and ion mass and therefore the measured eV/q spectrum can be converted to a u/q spectrum.
Under the assumption of singular charged ions, the u/q spectrum can be interpreted as an u spectrum.

In environments such as Titan's ionosphere where the ion temperatures are low, the expected peak flux for an ion species $\alpha$, occurs at energy $E_{\alpha}$ as approximated by equation (\ref{eq:energy}) \citep{Crary2009,Mandt2012,Westlake2014},

\begin{equation}
    E_{\alpha} = \frac{1}{2}m_{\alpha}(v_{sc} + v_{wind})^{2} + e\Phi_{sc} + 8kT.
    \label{eq:energy}
\end{equation}

Equation (\ref{eq:energy}) takes into account various effects that alter the peak energy flux such as ion temperature (8kT), spacecraft potential (e$\Phi_{sc}$) and ion wind ($v_{wind}$).
Through the inversion of equation (\ref{eq:energy}) we can find a mass associated with each peak in the energy spectra. The width of each beam is roughly equal to the thermal velocity of the ions, as described by equation (\ref{eq:BeamWidth}),

\begin{equation}
    \frac{M}{\Delta M} \approx \frac{v_{sc} + v_{wind}}{\sqrt{\frac{2kT}{m_{\alpha}}}}.
    \label{eq:BeamWidth}
\end{equation}

Spacecraft charging can affect the energy/charge of the particles measured by CAPS sensors.
Spacecraft potential values are inferred from another Cassini instrument, the Radio and Plasma Wave Science (RPWS) Langmuir probe \citep{Gurnett2004}. 
A constant offset between the Langmuir probe and CAPS IBS of +0.2 $\pm$ 0.15 V has been found by previous studies \citep{Crary2009,Westlake2011,Mandt2012}. 
For this study interpolated Langmuir probe potentials were used with an applied offset of +0.2V.
Also, a fixed temperature of 150K was applied during the analysis.
This value is representative of the temperatures seen in Titan's ionosphere which are between 100 and 200K \citep{Crary2009}.

Examining the magnitudes of the three terms in equation (\ref{eq:energy}) the dominant parameter is the ion wind. Taking the highest reported ion wind value of 260 ms\textsuperscript{-1}, this would represent a 9\% shift in the ion kinetic energy, i.e 2.8 eV shift for a 31.9 eV ion, which are the lightest ions examined in this study. The spacecraft potential term typically is between -0.5 and -2 eV \citep{Crary2009}. 2 eV represents a 6\% energy shift for a 31.9 eV ion. Lastly, an ion temperature of 150K corresponds to a 0.1 eV shift, a 0.3\% shift for a 31.9 eV ion. As seen in equation (\ref{eq:energy}), the ion temperature and spacecraft potential terms are not mass dependent, implying that for higher mass ions the energy shifts due to the ion temperature and spacecraft potential terms decrease as a proportion of the measured energy $E_{\alpha}$.

The ion wind calculation relies on previously identified major peaks.
The determination for which peaks to use was influenced by the work of \citet{Mandt2012}, who used the INMS instrument to calculate ion densities.
The three major peaks used in this study were the 28, 39 and 91 u/q peaks, these ions were used as they were found to be the most abundant ions within their respective mass groups \citep{Mandt2012}.

These peaks can be used to estimate the ion winds that are present in Titan's upper atmosphere and ionosphere.
The winds affect the detected energy of the ions by shifting the energy at which they peak in each spectrum as seen in equation (\ref{eq:energy}).
To construct mass spectra from the ion data this effect must be accounted for.
Using the previously stated abundant ions we can calculate the ion wind by using a ``mass-dependent effective spacecraft potential'' defined as \citep{Crary2009},

\begin{equation}
    \Phi^{'} \approx \Phi + \frac{mv_{sc}}{e}v_{wind},
    \label{eq:EffectivePotential}
\end{equation}

and by differentiating we find

\begin{equation}
    v_{wind} = \frac{e}{v_{sc}}\frac{d\Phi^{'}}{dm}  ,
    \label{eq:IonWindSpeed}
\end{equation}

where $\Phi$ is the spacecraft potential, $\Phi^{'}$ is the effective spacecraft potential,  $v_{sc}$ is the spacecraft velocity and $v_{wind}$ is the along track ion wind velocity.
By finding the effective spacecraft potential corrections for the three major mass peaks to be at their observed energies and calculating the gradient, the ion wind can be found using equation (\ref{eq:IonWindSpeed}).

\subsection{High Mass Methodology and Uncertainties} \label{sec:highmassmethodology}

\begin{figure*}
    \plotone{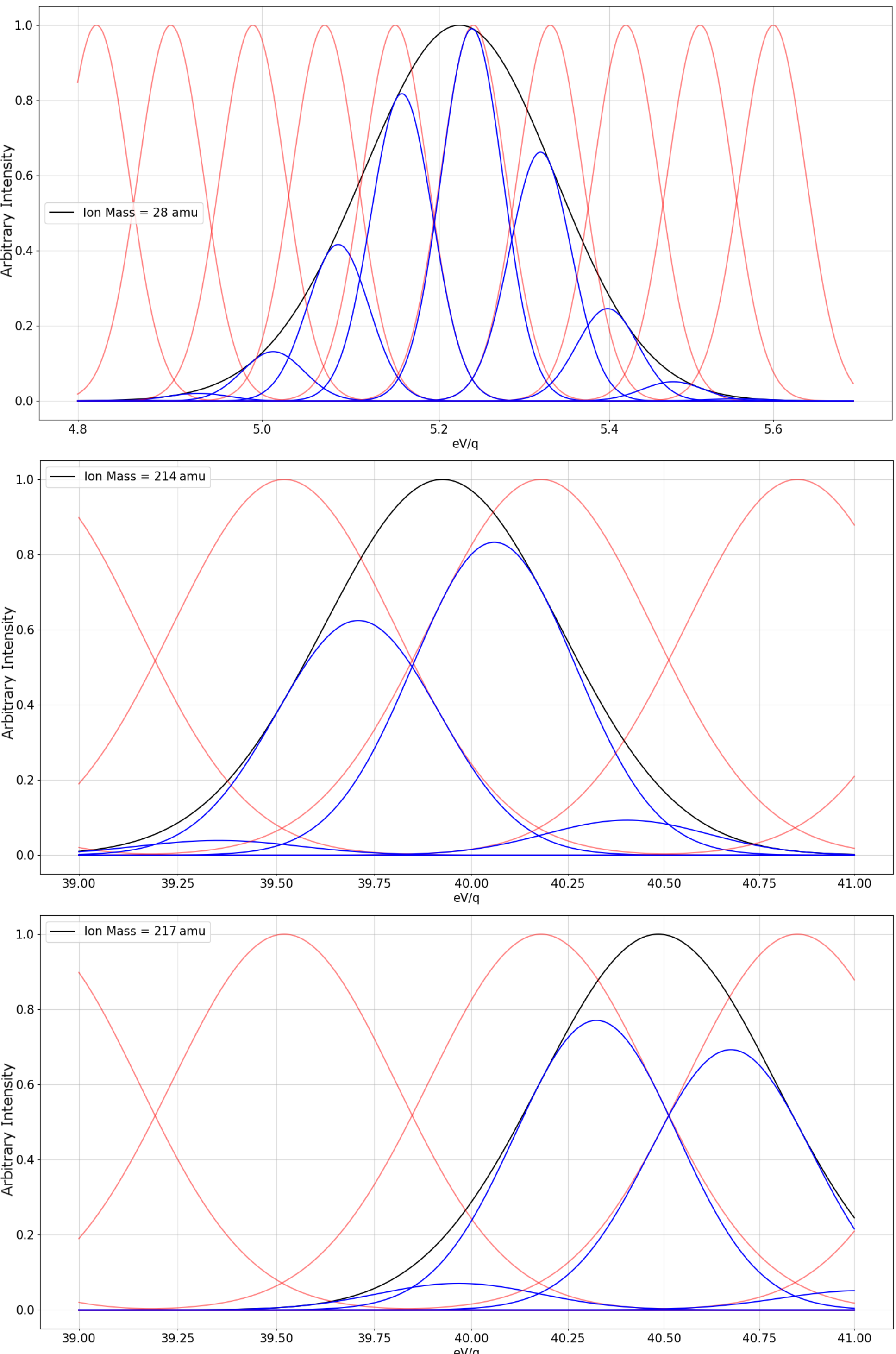}
    \caption{These three panels are used to illustrate the difference between observing low mass and high mass ion beams with the IBS sensor. For all three panels, the red lines represent the IBS response, with each gaussian curve representing a different IBS energy bin. The black line represents an observed ion beam, idealized with a gaussian and the width being determined by equation \ref{eq:BeamWidth}. The blue lines are each of the IBS energy bins (red) multiplied by the ion beam (black) and are representative of the spectra generated by IBS.}
    \label{fig:IBS_response}
\end{figure*}

A full analysis of the ions would require convolutions of the ion distributions at separate masses with the IBS energy response and then a subsequent fit to the observed data.
At the high masses studied here this becomes impractical as a result of the limited ability to resolve different masses due to the $\Delta E/E$ energy resolution at Full Width Half Maximum (FWHM) of the IBS sensor.
IBS was designed with a $\Delta E/E$ of 0.014 \citep{Young2004} and calibration tests measured a $\Delta E/E$ of 0.02 \citep{Vilppola2001}.   
Here we use a $\Delta E/E$ of 0.017, which was the operational $\Delta E/E$ value \citep{Young2004} and was used in \citet{Crary2009}.
A $\Delta E/E$ of 0.017 results in the effective $M/\Delta M$ mass resolution being limited to \textless60 by the instrument's energy response.

How the energy resolution acts as a limiting factor is shown in Figure \ref{fig:IBS_response}.
The top panel is representative of a low mass ion beam as observed by IBS, while the middle and lower panels show high mass ion beams observed by IBS.
As can be seen in the top panel, the low mass ion beam can be well resolved across multiple energy bins.
The sequences of blue peaks in Figure \ref{fig:IBS_response} can be compared to the first peak in Figure \ref{fig:example_spectra}, as both represent beams with similar ion masses.
The peak in Figure \ref{fig:example_spectra} appears across more bins due to the logarithmic scaling.

In contrast, the middle and lower panels show beams relating to ion masses of 214 and 217 u/q, both of which could cause a peak in the same IBS energy bin.
From this it could be concluded that the same methodology can be applied to both low and high masses but comparing with Figure \ref{fig:example_spectra} we can see this is not the case.
At the high energies around 40 and 50 eV/q, the beams are not resolved over three or four bins, with some peaks displaying similar intensity on either side of the peak.
This means that the fitting cannot be performed over several peaks and we are left to conclude that any ion mass that generates an ion beam with a peak energy within the FWHM of the energy bin is plausible.
Returning to the middle and bottom panels of Figure \ref{fig:IBS_response}, this means any mass between 214 and 217 u/q can generate a peak in the energy bin centred around 40.2 eV/q, thus creating the uncertainty in our mass resolution.

After the peaks in the energy spectra were found and converted to mass spectra, the masses were binned at a 1 u/q resolution, similar to that done by \citet{Crary2009}. 
This process does not generate a mass spectrum where one can analyze fragmentation patterns but shows which cations are more abundant than cations of neighboring masses.

The total mass uncertainty is calculated by adding in quadrature the uncertainty from the energy resolution and the uncertainty resulting from the 1 u/q binning process.
The energy resolution dominates the total uncertainty, being $\pm$ 2 or $\pm$ 3 u/q compared to $\pm$ 0.5 u/q from the binning.

\subsection{Studied Flybys} \label{sec:flybys}

The five flybys studied all occurred in 2009, all with similar characteristics such as Solar Zenith Angle (SZA), latitude and longitude as can be seen in Table \ref{table:TitanFlybyTable}.
CAPS was at a fixed actuation during these five flybys, meaning that rammed ionospheric ions were measured across the entire flyby, resulting in a greater number of data points than the previous actuating flybys.
The times when Cassini exited Titan's shadow can also be seen in the table.
In total 448 energy sweeps were used covering altitudes from 955 to 1001 km.
There is variation in the solar illumination conditions between the data used from the flybys.
The data selected from T55 was almost entirely in Titan's shadow, while T59 was illuminated the entire time with the other flybys containing a mixture.


\begin{table*}
\movetableright=-1in

\begin{tabular}{m{20pt}m{75pt}m{60pt}m{40pt}m{55pt}m{55pt}m{25pt}m{25pt}m{50pt}m{50pt}}
Flyby & UTC times & Titan shadow exit UTC time & Number of sweeps & Min altitude (km) & Max altitude (km) & Lat($^{\circ}$) & Long($^{\circ}$) & SZA Range ($^{\circ}$) & Titan Local Time \\ 
\hline
T55 & 21:25:13 - 21:28:08 & 21:28:06 & 88 & 966 & 1001 & -19.1 & 176.1 & 134-149 & 22:00:27 \\ 
T56 & 19:58:37 - 20:01:21 & 20:00:19 & 83 & 967 & 1000 & -28.0 & 175.4 & 127-143 & 22:00:57 \\
T57 & 18:30:57 - 18:34:09 & 18:31:37 & 97 & 955 & 1000 & -34.9 & 173.1 & 119-137 & 22:07:31 \\
T58 & 17:02:39 - 17:05:25 & 17:01:21 & 84 & 966 & 999 & -44.7 & 172.4 & 112-128 & 22:07:42 \\
T59 & 15:32:26 - 15:35:39 & 15:29:03 & 96  & 957 & 1000 & -54.7 & 171.6 & 103-120 & 22:08:20 \\
\end{tabular}

\caption{Table displaying the data used in this study. The time where Cassini emerged from Titan's geometric shadow is shown in the under "Titan shadow exit UTC time". Min and Max altitudes demonstrate that these samples were taken at some of the lowest regions of Titan's ionosphere traversed by Cassini. Latitude(Lat) and Longitude(Long) are displayed at the closest approach during the flyby. Solar Zenith Angle(SZA) Range is the range of SZA values observed in the sample range. Titan Local Time is shown in the last column.   \label{table:TitanFlybyTable}}
\end{table*}

\section{Results} \label{sec:results}

\begin{figure*}
    \centering
    \includegraphics[width=1\textwidth]{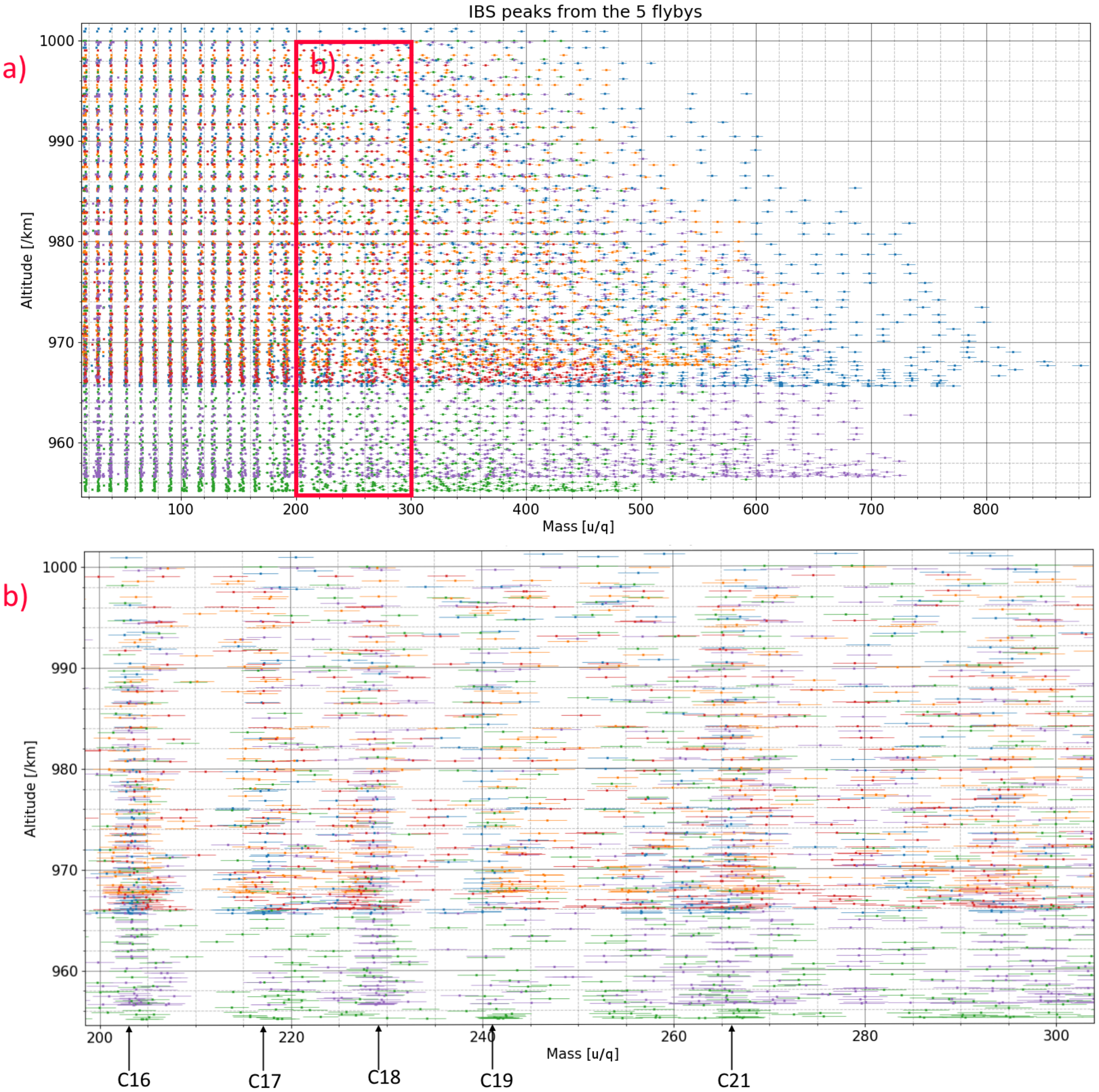}
    \caption{a) An altitude-mass plot for every peak identified by the peak finding algorithm at the altitudes examined. The error bars represent the instrument uncertainty for the peak detection. Each color represents a different flyby, T55 - blue, T56 - orange, T57 - green, T58 - red, T59 - purple. Panel b) is a section of panel a) between 200 and 300 u/q. Clear grouping of ions can be seen, notably around 203, 217, 229, 241 and 266 u/q.}
    \label{fig:IBS_altitudes_zoom}
\end{figure*}

Through the use of the previously described algorithm across the 448 CAPS IBS energy sweeps, the occurrence of mass groups and the group peaks can be studied.
Shown in Figure \ref{fig:IBS_altitudes_zoom}a is every peak identified across the five flybys, plotted against altitude.
The previously found trend of heavier ions occurring at lower altitudes \citep{Crary2009} can be seen.
Figure \ref{fig:IBS_altitudes_zoom}b displays the peaks between 200 and 300 u/q and below 1000km.
With this plot several groups can be seen with distinct ion clustering around 203, 217, 228, 241 and 266 u/q.
The T57 and T59 flybys have closest approaches of 955 and 957 km, around 10 km lower than the other three flybys that can be seen in the figure.

The low mass ions from 0-100 u/q have been well studied by both the CAPS and INMS instruments \citep{Mandt2012,Westlake2014}.
Groups of intermediate mass (100-200 u/q) positive ions were found along with major peaks with a 12-14 u/q spacing between peaks.
This study focuses on the high masses between 170 and 310 u/q outside the mass range of the INMS instrument.
The structure is less clearly defined than the lower mass ranges and therefore only mass groups and significant peaks are focused on in this study.

\subsection{Previously Reported Ion Groups (100 - 200 u/q)} \label{sec:results-lowmass}

\begin{figure*}
    \includegraphics[width=1\textwidth]{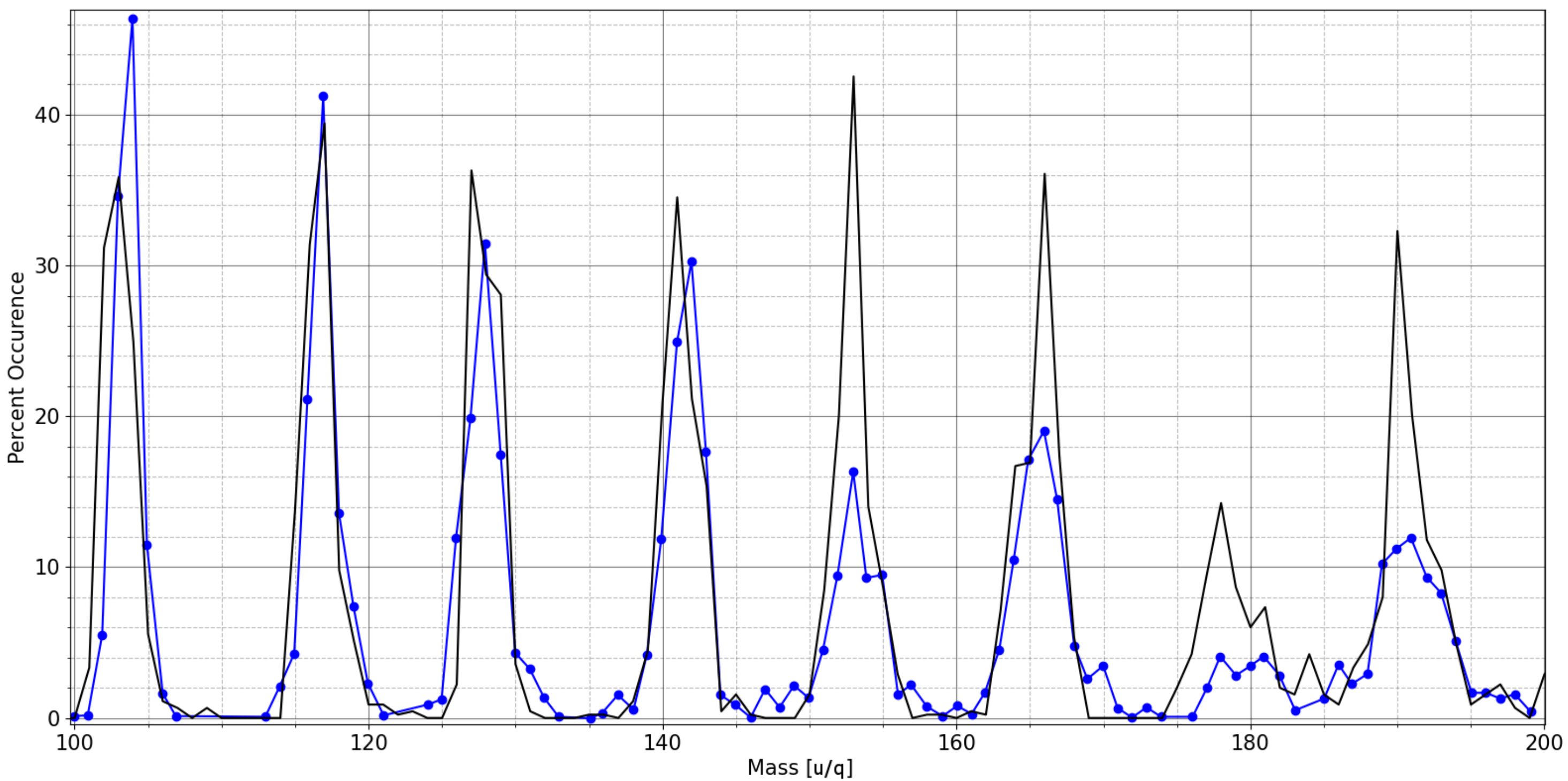}
    \caption{The black line represents the summed percent occurrence of the peaks over the flybys with 1 u/q binning from the present study. The blue line represents the percent occurrence from \citet{Crary2009}. Good agreement can be seen, with the present study showing more defined peaks for the ion groups around 180 and 190 u/q.}
    \label{fig:crarycomparison}
\end{figure*}

For the comparison to \citet{Crary2009}, after the peaks were found in the energy spectra and converted to mass values, the mass values were binned at a 1 u/q resolution. 
Then the number of occurrences of a peak in each 1 u/q bin across the five flybys was summed and then divided by the total number of sweeps examined across the flybys. 
This generates a percent occurrence value, representing how often a peak occurred in the studied data and is shown by the black line in Figure \ref{fig:crarycomparison}.
Eight mass groups were found in the 100 to 200 u/q mass range with each group being centered around a significant peak.
For comparison, the percent occurrence found by \citet{Crary2009} is shown by the blue line.

This comparison is necessary due to this study applying different methodology than that applied previously in \citet{Crary2009}.
The main difference is that the cross-calibration between INMS and IBS data was not performed in this study, but was applied in the previous \citet{Crary2009} study.
The comparison helps to link the existing study to the higher mass resolution measurements of INMS. 
Other differences include the mass peaks found by \citet{Crary2009} being derived from interpolating the energy spectra to 1 u/q mass bins using fit determined parameters and then identifying peaks.
Lastly, \citet{Crary2009} studied flybys up to January 2008 and this study examines flybys during 2009, therefore, a comparison at this mass range is useful to check for variation with time.

Although there are methodology differences this study reproduces the mass peaks identified by \citet{Crary2009} fairly accurately.
Agreement between the new methodology and previous work gives us confidence that the  cross-calibration is not necessary for this study and to extend the method to study high mass ions above 200 u/q.

\begin{figure*}
    \includegraphics[width=1\textwidth]{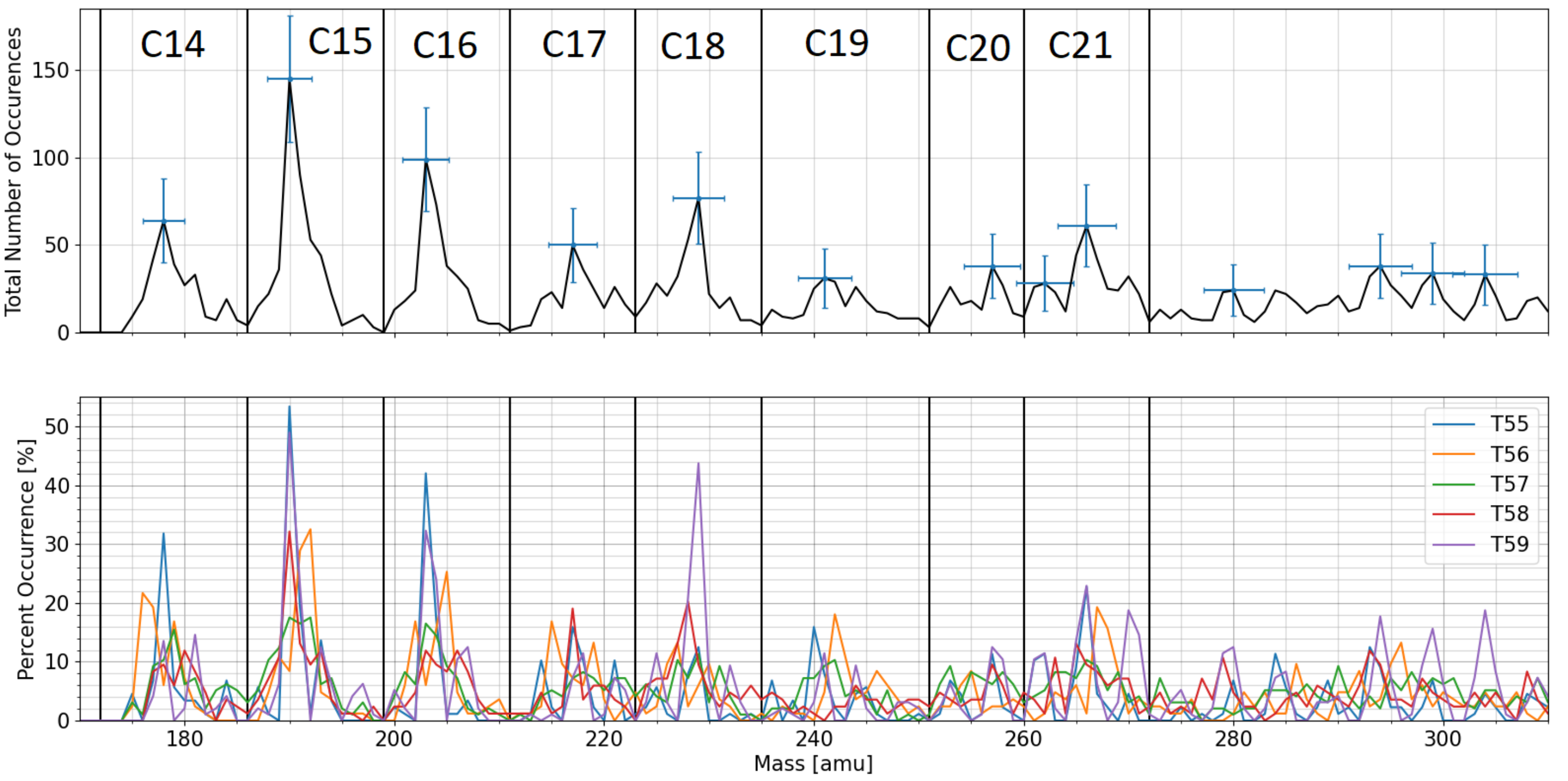}
    
    \caption{The top panel displays the total number of times a peak occurs in each 1 u/q bin.
    Vertical lines indicate the boundaries between the different ion groups, with the boundaries being defined by local minima in the total number of occurrences.
    Each group is labeled with C\#, where \# indicates the number of heavy (carbon/nitrogen/oxygen) ions present.
    The bottom panel represents the percentage occurrence by flyby, with the same group definitions as the top panel.
    The percentage occurrence is by flyby, for example the large peak for the T59 flyby in the C18 group occurs in 44\% of the 96 sweeps examined.
    The vertical error bars in the top panel shown are three times the standard error relating to Poisson counting statistics, meaning there is a 99\% probability that the number of occurrences lies within this range.
    The horizontal error bars are a combination of the uncertainty resulting from the energy resolution and the uncertainty from a 1 u/q binning. 
}
    \label{fig:highmass_spectra}
\end{figure*}

\begin{figure*}
    \includegraphics[width=1\textwidth]{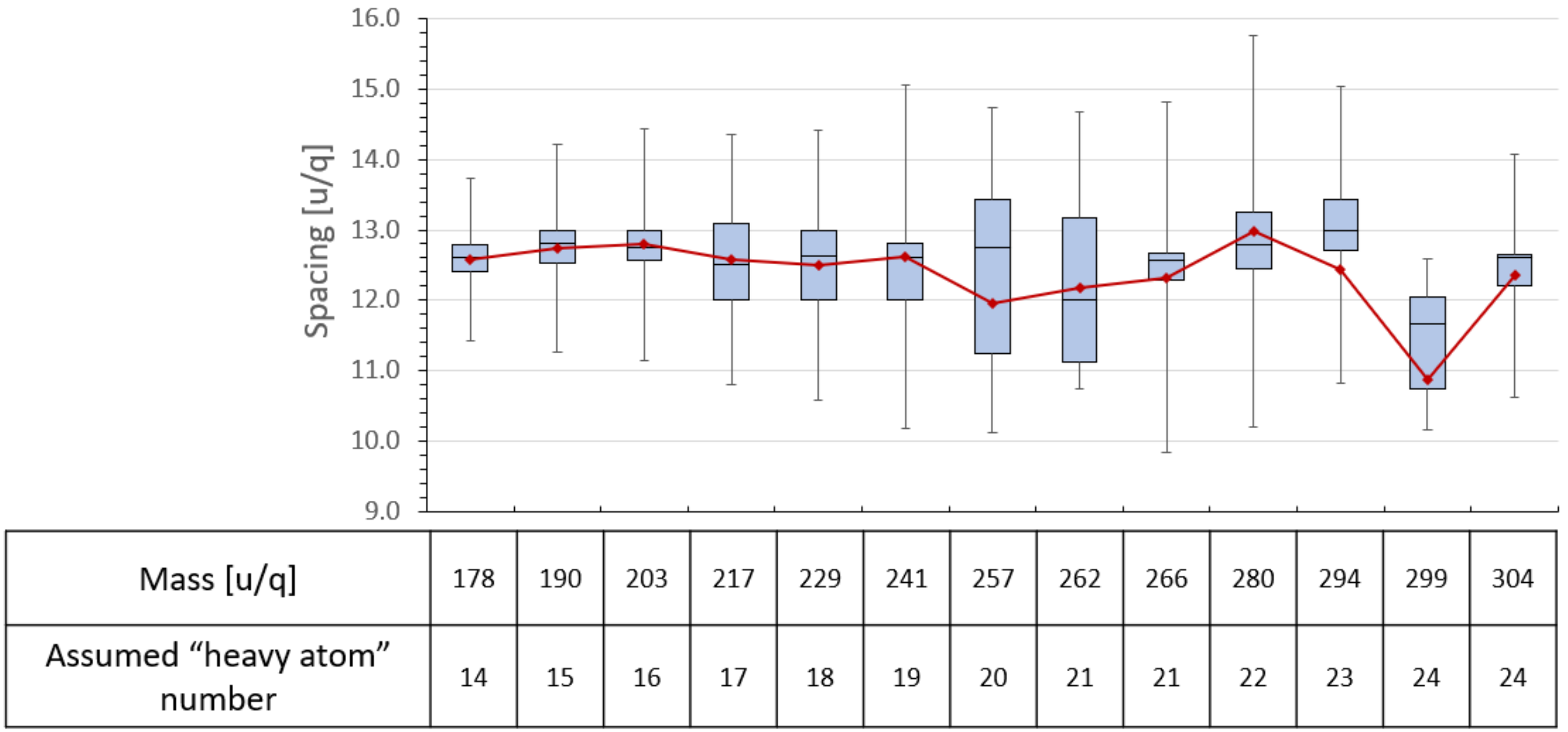}
    \caption{A box and whisker plot demonstrating the mass spacing between the significant observed peaks under the assumption of a number of ``heavy atom'' (carbon/nitrogen/oxygen) in a molecular ion. This is calculated by finding the mass difference between pairs of peaks and dividing through by the difference in ``heavy atoms''. The mean spacing is plotted in red and is largely in the 12 and 13 u/q range. The box shown represents the interquartile range of the peak spacing along with the median. The low mean spacing of 11 u/q for the 299 u/q peak is due to the close neighboring masses at 294 and 304 u/q, a similar effect can be seen for the 257 and 262 u/q peaks.
}
    \label{fig:peakspacing}
\end{figure*}

\subsection{High Mass Ions (170 - 310 u/q)} \label{sec:results-highmass}

The high mass ions and the mass group structure is displayed in Figure \ref{fig:highmass_spectra}.
The mass group numbering (C\#) indicates the number of heavy atoms (carbon/nitrogen/oxygen) present.
Clear structure can be seen for the C14 to C19 groups, corresponding to masses between 172 and 251 u/q.
These groups have a major peak and well-defined troughs between the groups.
Furthermore, below 251 u/q the spacing between peaks of 12 to 14 u/q agrees with the trend observed at lower masses.

The C20 and C21 groups have major peaks but also display minor peaks and the troughs between groups are not as evident. 
At these masses, the 12 to 14 u/q spacing trend breaks for C20, with a 16 u/q gap between C19 and C20 and a 9 u/q gap between C20 and the C21 major peaks.
However as there are uncertainties of $\pm$ 3 u/q due to the energy resolution of IBS at these masses, it is plausible that the spacing is closer to 12 or 14 u/q.
Alternatively, there is a minor peak in the C20 group at 253 u/q, which would correspond to 12 and 13 u/q gaps between the C20 peak and its neighboring groups. 

Above 270 u/q no clear peak and trough structure can be seen in Figure \ref{fig:highmass_spectra}.
Several significant peaks do exist, one such peak is seen at 280 u/q that would be associated with an ion containing 22 heavy atoms.
At the highest masses studied here peaks are seen at 294 $\pm$ 3, 299 $\pm$ 3 and 304 $\pm$ 3 u/q.
These peaks however do not have a distinct gap between them due to the uncertainties in measurement.
Given the range of masses, it is likely there are multiple peaks and these are likely associated with ions containing 23 or 24 heavy atoms.


\subsubsection{Peak Spacing} \label{sec:results-peakspacing}

Further analysis of the mass spacing between peaks was performed to aid the understanding of growth pathways.
This was done by finding the mass difference between each possible pair of peaks and dividing each mass difference by the number of assumed heavy atoms, for example, the mass difference for the 178 and 217 u/q peaks is 39 u/q and the difference in assumed heavy atoms is 3, generating a peak spacing of 13 u/q. 
This is repeated for all possible peak pairs and then the peak spacing is averaged. 
The results from this can be seen as a box and whisker plot in Figure \ref{fig:peakspacing}, with the mean plotted in red.
The lower and upper bounds of the whiskers are due to the combined uncertainty from both peak measurements.
For example, the 203 and 217 u/q peaks both have an uncertainty of $\pm$ 2 u/q, meaning the peak spacing for this pair could be between 10 and 18 u/q.
The value shown for the lower bound is the average of this minimum spacing across all peak pairs, a similar calculation gives the value for the upper bound.
Some peaks have large interquartile ranges such as 257, 262 and 299 u/q, these are caused by neighboring peaks with small mass differences. 
These peaks could also be indicative of a molecule containing a heavier atom than carbon, such as nitrogen or oxygen, which would then affect the spacing between the peaks.

The mean spacing across all peaks is 12.4 u/q and the lower to upper quartile range typically covers the 12 to 13 u/q.
These masses would be consistent with a C or CH addition to the molecular ions.

\section{Discussion} \label{sec:Discussion}

Many structures have been hypothesized for the large molecular ions in Titan's ionosphere.
These have included polyacetylene, polyynes, cyanopolyynes \citep{Wilson2003, Lavvas2008a, Lavvas2008b, Vuitton2019}.
(HCN)\textsubscript{n} polymers have been proposed \citep{Wilson2003}, as well as HC\textsubscript{3}N/C\textsubscript{2}H\textsubscript{2} copolymers \citep{Lebonnis2002,Lavvas2008b}.
PAHs and PANHs have been proposed as well \citep{Lopez-Puertas2013,Zhao2018}, resulting from the detection of benzene in Titan's upper atmosphere \citep{Waite2007}.
Amines and imine polymers have been proposed to contribute to Titan's haze due to either the prerequisite monomers being found in Titan's atmosphere or the molecules being detected in tholin experiments \citep{Vuitton2006,Cable2014,Skouteris2015}.
Fullerenes have also been proposed to exist \citep{Sittler2020}.

For assessing the composition of the ion peaks, we compare the masses associated with the ion peaks, with the masses consistent with different structures.
This includes those structures mentioned above as well as other possible structures such as long aliphatic hydrocarbons with different levels of saturation.
These aliphatic hydrocarbon chains include alkanes, alkenes, dienes, trienes and alkynes.
A table summarizing the comparison of peaks to possible structures is shown in Figure \ref{fig:structurestable}.
Combinations of aliphatic hydrocarbons and various polymers can explain the observed ion peaks but the chemical structures that are consistent with all peaks are Polycyclic Aromatic Compounds (PACs).
Furthermore, chemical structures such as (HCN)\textsubscript{n} polymers for example, are consistent with peaks for n = 7, 8, 9 and 11, but are not present for n = 10, which is unlikely to occur.
This does not rule out the presence of aliphatic compounds but rather that the most abundant ions observed in this study are consistent with PACs.
PACs represent a combination of PAH and/or polycyclic aromatic heterocycles containing nitrogen.
Oxygen-bearing polycyclic aromatic molecules are also possible. 
There are also likely methylene substituted derivatives of these compounds, as proposed in \citet{Ali2015}.

PACs possibly have a role in the formation of aerosols \citep{Lavvas2011}.
Early experiments found that Titan tholin analogs contain PAHs \citep{Sagan1993} and also that PAHs may adsorb onto the tholins.
Later studies have also found PANHs and PAHs in Titan tholin analogs \citep{Trainer2013,Yoon2014,Mahjoub2016,SciammaOBrien2017}.
\citet{Gautier2017} discuss how aerosol growth is affected depending on the number of aromatic rings in the reactant composition, finding that benzene and pyridine caused production of co-polymeric structures while double-ringed aromatic produced quasi-pure polymeric structures.
Other studies have also investigated the impact of ions interacting with charged aerosols, demonstrating that they contribute to the growth of these particles \citep{Lavvas2013}.
These findings indicate that the composition of Titan's aerosols is likely affected by the composition of the positive PAC ions studied here.

\citet{Westlake2014} examined three hypotheses for the production of high mass ions in Titan's ionosphere.
They concluded that the most viable method for creating the large molecules were through ion-molecule reactions with neutral acetylene, ethylene and hydrogen cyanide.
\citet{Ali2015} further investigated the cation, anion and neutral chemistry that could be present in Titan's ionosphere through ion-molecule reactions.
Low-temperature formation pathways of PAHs through radical-neutral reactions have been shown to exist for molecules with up to four six-membered rings \citep{Zhao2018}.


Comparing the findings presented here with the ion-molecule reactions proposed in \citet{Westlake2014} and \citet{Ali2015} we find the ion group peak masses agree with the expected masses resulting from ion-molecule reactions.
Furthermore, comparing with proposed neutral PAHs/PANHs from \citet{Lopez-Puertas2013}, we find the reported neutral molecules have comparable masses, suggesting ionization and electron recombination are a factor.
These two comparisons suggest a coupling between the cationic and neutral phases at the studied mass range.

\begin{longrotatetable}
\movetabledown=10mm
\begin{figure*}
    \includegraphics[width=1.30\textwidth]{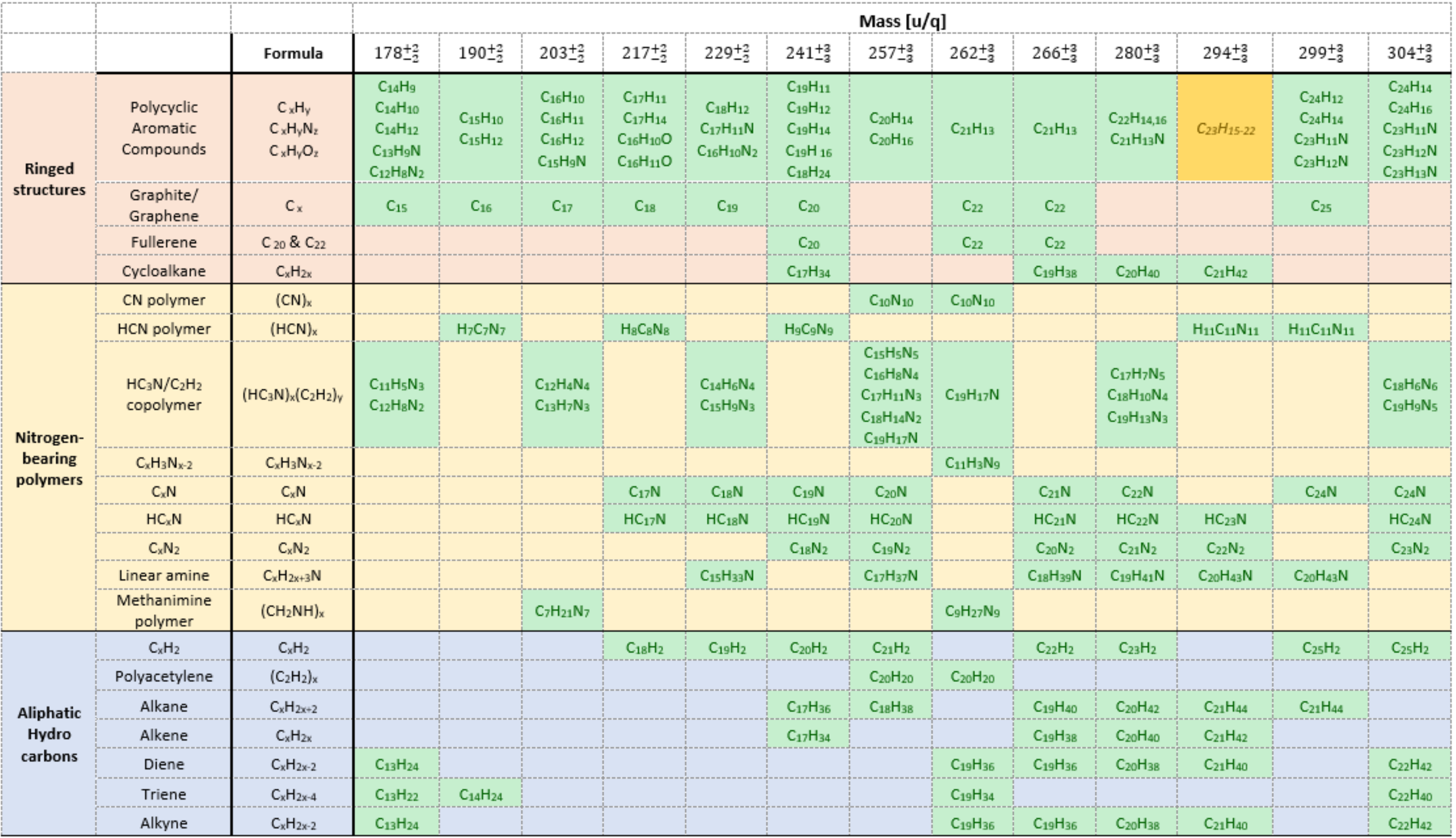}
    \caption{A table showing what structures are consistent with significant peaks. The structures are broken into three groups and highlighted in different colors for clarity: ringed structures containing aromatic and cyclic molecules, nitrogen-bearing aliphatic molecules and aliphatic hydrocarbon molecules.  Any molecule that lies within the mass uncertainty associated with a peak is considered plausible. Polycyclic Aromatic Compounds are consistent with all observed ion peaks, followed by a graphite- or graphene- like structure, consistent with nine of the thirteen observed peaks. The list of polycyclic aromatic molecules considered are from the NASA Ames PAH IR database (except for the 294 u/q peak) and should be viewed as not-exhaustive.}
    \label{fig:structurestable}
\end{figure*}
\end{longrotatetable}

\subsection{Ion Mass Group Composition} \label{sec:Discussion-composition}

Here we discuss the composition of the positive ions and a possible growth mechanism.
The C14 and C15 group peaks have masses consistent with cationic three ringed PACs.
Examining the C14 group across all flybys, the major peak lies between 176 and 181 u/q in the IBS data.
\citet{Ali2015} suggested hydrocarbon cations of C\textsubscript{14}H\textsubscript{9}\textsuperscript{+}, C\textsubscript{14}H\textsubscript{11}\textsuperscript{+} and C\textsubscript{14}H\textsubscript{13}\textsuperscript{+} as well as nitrogen-bearing C\textsubscript{13}H\textsubscript{10}N\textsuperscript{+} and C\textsubscript{13}H\textsubscript{12}N\textsuperscript{+} ions. 
These molecules lie within the 176 and 181 u/q range given the associated errors, meaning we cannot distinguish between the PAHs and nitrogen-containing PACs.
Similarly, the cations in the C16, C17 and C18 groups can be associated with four ringed PAC cations and the C19, C20 and C21 with five ringed PAC cations.


The C15 group frequently occurs in all flybys generating a strong peak at 190 $\pm$ 2 u/q.
\citet{Ali2015} describe the mechanism of substituting  CH\textsubscript{2}\textsuperscript{+} in place of a hydrogen atom in a neutral PAC to generate cations, this would increase the mass of the ion by 13 u/q, consistent with our peak spacing finding of 12 to 13 u/q.
Subsequent reactions with neutral ethylene creates larger cations alongside ejection of one or two H\textsubscript{2} molecules.
A similar mechanism is also proposed by \citet{Westlake2011}, who investigated both ion-molecule reactions with neutral acetylene and ethylene. 

The same CH\textsubscript{2}\textsuperscript{+} substitution mechanism can be applied to generate all observed ion group peaks.
From Figure \ref{fig:structurestable} for example, possible growth chains of CH\textsubscript{2}\textsuperscript{+} substitutions could be C\textsubscript{14}H\textsubscript{9} $\rightarrow$ C\textsubscript{15}H\textsubscript{10}\textsuperscript{+} followed by C\textsubscript{15}H\textsubscript{10} $\rightarrow$ C\textsubscript{16}H\textsubscript{11}\textsuperscript{+}.
These chains could also exist at the upper end of the studied mass range as well, C\textsubscript{20}H\textsubscript{14} $\rightarrow$ C\textsubscript{21}H\textsubscript{13}\textsuperscript{+} alongside a H\textsubscript{2} ejection followed by C\textsubscript{21}H\textsubscript{13} $\rightarrow$ C\textsubscript{22}H\textsubscript{14}\textsuperscript{+}.
These chains could similarly occur for nitrogen- or oxygen- bearing PACs.  
Although this growth pathway is consistent with the observed peak spacing it does not rule out other reactions occurring with PAC cations.

Oxygen ions entering the ionosphere and the presence of H\textsubscript{2}O, CO\textsubscript{2} and CO \citep{Cui2009composition,Horst2012} imply that oxygen-bearing polycyclic aromatic molecules are plausible, although are likely to have a small contribution compared to PAHs or PANHs.
These oxygen ions charge transfer with neutrals generated thermal O atoms \citep{Vuitton2019} and interstellar medium experiments have found reaction pathways between atomic oxygen and the C\textsubscript{10}H\textsubscript{8}\textsuperscript{+} naphthalene cation \citep{LePage1999b}.
\citet{LePage1999a} studied reactions between various pyrene cations and atomic oxygen and found that C\textsubscript{16}H\textsubscript{9}O\textsuperscript{+} and C\textsubscript{16}H\textsubscript{10}O\textsuperscript{+} could be created.
The reaction with CO could be significant and has been shown to incorporate oxygen into tholins in atmospheric simulation experiments \citep{Fleury2014}. 
Other reactions involving OH, H\textsubscript{2}O and other oxygen-bearing molecules cannot be ruled out but are not explored here.
The formation pathways and reaction rates for oxygen-bearing polycyclic aromatic molecules are poorly understood but their presence cannot be ruled out by this study.



\subsection{Variations between Flybys} \label{sec:Discussion-percenvariation}

\begin{figure*}
    \includegraphics[width=1\textwidth]{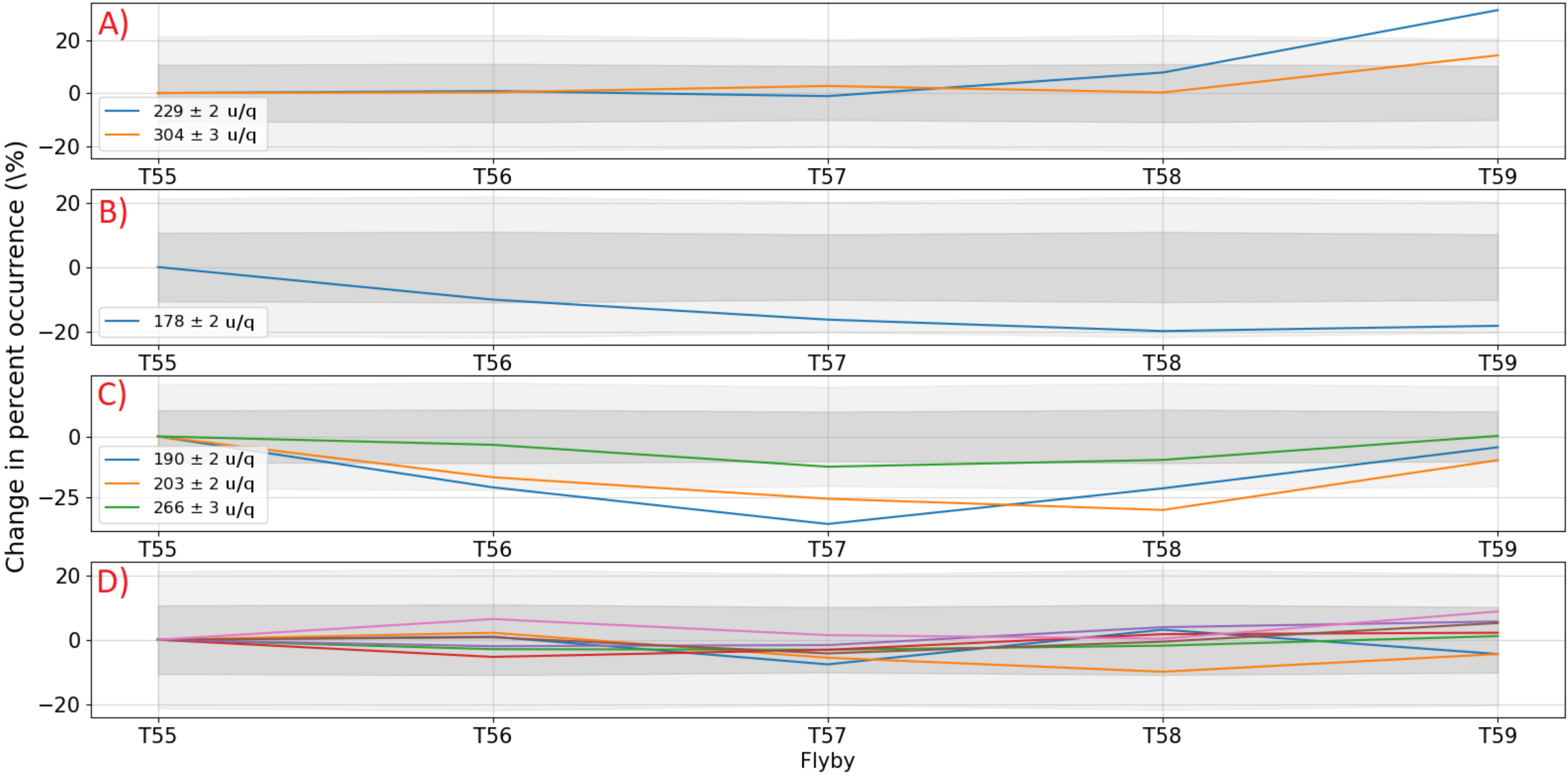}
    \caption{The change in percent occurrence of the observed peaks, as compared to T55. The uncertainties are Poisson counting statistics, with the dark grey and light grey regions representing one and two times the standard deviation respectively. The peaks are categorized into four groups, panel A) shows peaks which occurred more frequently under illuminated conditions. Conversely, panel B) shows a peak which occurred more frequently under unilluminated conditions. Panel C) shows the peaks which had a minimum during the flybys and panel D) are the remaining seven observed peaks that show no clear variation, with all points lying within one standard deviation.  
}
    \label{fig:percentvariation}
\end{figure*}

Examining the different characteristics for the five studied flybys: the closest approach altitudes were between 955 and 967 km, the closest approach longitudes were within a 5 degree range and the closest approach local times were all close to 22:00. 
However, the SZA values range from 103 to 149, which is primarily due to the flyby latitude increasing from -19$^{\circ}$ during T55, to -55$^{\circ}$ during T59.
There is also variation in the percentage occurrence of some observed peaks between different flybys.
This could be a signature of varying production, loss and transport rates of the ions associated with the peaks.
Furthermore, it could be linked to neutral and anion chemistry that occurs, which varies between the day and nightside of Titan \citep{Vuitton2019}.
Examining Figure \ref{fig:percentvariation}, six of the thirteen significant peaks display clear variations.
Some peaks occur more frequently with more solar radiation while one peak occurs less frequently. 
Furthermore, some peaks can be seen to occur less frequently in the transition between illuminated and unilluminated conditions.

From Figure \ref{fig:percentvariation}, we can see that the 229 $\pm$ 2 and 304 $\pm$ 3 u/q peaks occur more frequently in the T59 flyby.
Examining the percentage occurrence of the 229 u/q peak in Figure \ref{fig:highmass_spectra}, there is a 35\% increase for T59 compared to T55.
The trend for the 304 $\pm$ 3 u/q peak is not as clear, with the significant increase in occurrence occurring only during the T59 flyby.

In contrast, panel B) in Figure \ref{fig:percentvariation} shows the only example of a significant decrease in percent occurrence, associated with the 178 $\pm$ 2 u/q peak in the C14 ion mass group.
Between the dark T55 flyby and the illuminated T59 flyby the peak occurrence drops by around 20\%.
The C14 group is the least frequently seen group in the 100-200 u/q range in both this study and \citet{Crary2009}, which could be related to this observed variation.

The third panel in Figure \ref{fig:percentvariation} shows three peaks which have an occurrence minimum during the five studied flybys.
The 190 $\pm$ 2 and 266 $\pm$ 3 u/q peaks have a minimum percentage occurrence during the T57 flyby while 203 $\pm$ 2 u/q minimum is during T58. 


Due to Titan's extended atmosphere the effects of solar EUV radiation on ions are seen past the terminator, up to SZA values of 135$^{\circ}$ at 1000km \citep{Agren2009,Cui2009,Coates2009,Coates2011,Wellbrock2009,Wellbrock2013,Wellbrock2020}, although at high SZA values the solar radiation is attenuated by passing through more of Titan's atmosphere
\citet{Cui2009} studied the diurnal variations of positive ions, finding that light ions were strongly depleted on the nightside while the heavy ions were moderately depleted and the most significant diurnal variations were seen at low altitudes.
These depletions are linked to the loss of ion production from solar EUV radiation.

Similar to the light cations, light anions (\textless 50 u/q) have been shown to have higher densities on the dayside \citep{Wellbrock2012,Wellbrock2020,Mihailescu2020}.
On the other hand, \citet{Wellbrock2009,Wellbrock2020} and \citet{Coates2009,Coates2011} show that heavy anion densities under illuminated conditions past the terminator are lower than unilluminated conditions in the deeper nightside ionosphere, as observed during flybys such as T57.
The higher densities of heavy anions in unilluminated conditions could provide a competing pathway to heavy cation production.
This could be through higher anion production rates or through lower anion loss rates, an example of a loss process would be photodetachment.
Photodetachment efficiency increases with ion size \citep{Lavvas2013,Wellbrock2019}, which would remove heavy anions from the dayside allowing for higher heavy cation densities.
Ionization energy is the energy needed to remove an electron from a molecule while electron affinity is the amount of energy released when an electron is added to a molecule.
There is a general trend of increasing electron affinity and decreasing ionization energy with increasing PAH mass \citep{christodoulides1984electron,Dabestani1999}, meaning larger PAHs can more readily form positive or negative ions under favorable conditions.

\citet{Wellbrock2013} identified several mass groups of negative ions, with the 190-625 u/q mass group displaying the highest density of all groups at an average peak altitude of 1000km.
The flybys in this study are at a similar altitude and the mass range studied covers this negative ion group.

Due to the complex cation-neutral-anion chemistry and lack of information on reaction rates at these high masses, we cannot distinguish between different factors such as anion chemistry, ion transport and changing ion-neutral pathways.
Future investigations that examine flybys in the noon/midnight regions would help identify photochemical effects.
Furthermore, future chemical models could aid identification of these heavy molecular ions through examining recombination rates and ion-neutral reactions to see which ions are consistent with observed variations.

\subsection{Comparison with Neutral PACs in Titan's Atmosphere} \label{sec:Discussion-neutralcomparison}

\citet{Lopez-Puertas2013} studied an IR emission from Titan and applied a fitting algorithm to identify neutral PACs in the upper atmosphere.
7 of their 19 most abundant PACs lie in the 170-310 u/q mass range.
These seven molecules are
C\textsubscript{13}H\textsubscript{9}N  (179 u/q), C\textsubscript{12}H\textsubscript{8}N\textsubscript{2} (180 u/q),
C\textsubscript{14}H\textsubscript{16} (184 u/q), C\textsubscript{16}H\textsubscript{10}N\textsubscript{2}  (230 u/q), C\textsubscript{20}H\textsubscript{10} (250 u/q), C\textsubscript{20}H\textsubscript{14} (254 u/q) and C\textsubscript{22}H\textsubscript{16} (280 u/q).
These neutral molecules are in the even numbered mass groups: C14, C18, C20 and C22.

Three of these molecules are in the C14 group, C\textsubscript{13}H\textsubscript{9}N, C\textsubscript{12}H\textsubscript{8}N\textsubscript{2} and C\textsubscript{14}H\textsubscript{16}.
The first two of these neutrals could be related to the major C14 peak in the IBS data at 178 u/q. 
\citet{Lopez-Puertas2013} reported C\textsubscript{14}H\textsubscript{16} to be present at a low concentration and in this study we find a minor peak (percentage occurrence below 10\%) at 184 $\pm$ 2 u/q that occurs in four of the studied flybys. 
This peak could result from the ionization of the reported C\textsubscript{14}H\textsubscript{16} neutral.

 C\textsubscript{16}H\textsubscript{10}N\textsubscript{2} is in the C18 group and could be related to the ion mass peak at 229 u/q. 
The two molecules in the C20 group, C\textsubscript{20}H\textsubscript{10} and C\textsubscript{20}H\textsubscript{14}, were some of the least abundant molecules reported by \citet{Lopez-Puertas2013}.
The major peak in the C20 group is at 257 $\pm$ 3 u/q with a minor peak at 253 $\pm$ 3 u/q.
This puts both molecules within the error of the ion peak.
However, there is no clear peak structure at 250 u/q with the region between 248 and 251 u/q being a trough between the C20 and C21 groups.
This could indicate that we see a higher level of hydrogenation in the cationic phase.

Although we do not find clear structure for the C22 group, there is a peak at 280 $\pm$ 3 u/q therefore a similar mass to the neutral C\textsubscript{22}H\textsubscript{16} at 280 u/q.
This peak strongly occurs during the T58 and T59 flybys, peaking at 279 u/q and 280 u/q respectively.

In summary, from the comparison between the ionic and neutral phases, the masses of previously observed neutral PACs in Titan's atmosphere correlate with cations in the ionosphere.
The only exception found was the neutral PAH C\textsubscript{20}H\textsubscript{10}.
The comparisons between the observed abundant neutral molecules and the frequently occurring ion peaks indicate that there is a need for combined study of the neutral and cationic phases to fully characterize Titan's ionosphere.
In addition, further study of what neutral PAHs exist in  Titan's ionosphere could aid with interpretation in future ion compositional studies.

\section{Conclusions} \label{sec:Conclusions}

The composition of positive ions between 170 and 310 u/q has been examined during five CAPS actuator-fixed flybys that took place during 2009. 
The composition was investigated by examining how frequently a peak occurs at a given mass.
Up to 275 u/q there is a clear ion group structure of peaks and troughs, where the peaks are ion masses with a high relative abundance.
Conversely, the troughs are associated with ion of low relative abundance.
Between 275 and 310 u/q, several prominent peaks are found but with no clearly associated group structure.
The identified ion peaks are consistent with various aliphatic compounds, however, polycylic aromatic compounds, such as PAHs/PANHs, are found to be consistent with all identified peaks.
The spacing between peaks was found to be between 12 and 13 u/q, with an average spacing of 12.4 u/q consistent with the addition of C or CH.

Variation in the occurrence of prominent peaks indicates changes in the photochemistry, most likely due to differing solar radiation conditions, at the studied 170-310 u/q mass range.
Comparisons with neutral PAHs/PANHs findings in the atmosphere reveal that the newly found ion peaks are found at similar masses to the abundant molecules in the neutral phase.
This implies coupling between the ion and neutral phases at these masses.

These are the largest distinct positive ion groups reported so far in Titan's ionosphere.
The discovery of these groups will aid future atmospheric chemical models of Titan through the identification of what heavy positive ions are prominent, aiding the understanding of which reaction pathways are plausible to create the previously found heavy negative ions and neutrals.
The future Dragonfly mission to Titan will study the surface composition and prebiotic chemistry.
Given that the complex organic compounds created in Titan's atmosphere form the particles that comprise the haze found at lower altitudes and that these particles are expected to fall onto the surface, understanding the origin of these organic compounds gives helpful insights into the interpretation of future results.

\acknowledgments

R. P. Haythornthwaite is supported by Science and Technology Facilities Council (STFC) studentship 2062537. 
A. J. Coates, G. H. Jones and A. Wellbrock acknowledge support from the STFC consolidated grant to UCL‐MSSL ST/S000240/1.
We acknowledge funding from ESA via the UK Space Agency for the Cassini CAPS Operations team. 
Work in the USA was supported by Southwest Research Institute internal funding.
Cassini CAPS data are available from the \dataset[NASA PDS service]{https://pds-ppi.igpp.ucla.edu/mission/Cassini-Huygens}.
We thank the reviewers for their helpful comments that have improved the paper.

\software{astropy \citep{2013A&A...558A..33A,2018AJ....156..123A}, matplotlib \citep{matplotlib}, NumPy \citep{2011CSE....13b..22V}, SciPy \citep{2020SciPy-NMeth}}




\vspace{32mm}

\bibliography{references}{}

\begin{thebibliography}{}
\expandafter\ifx\csname natexlab\endcsname\relax\def\natexlab#1{#1}\fi
\providecommand{\url}[1]{\href{#1}{#1}}
\providecommand{\dodoi}[1]{doi:~\href{http://doi.org/#1}{\nolinkurl{#1}}}
\providecommand{\doeprint}[1]{\href{http://ascl.net/#1}{\nolinkurl{http://ascl.net/#1}}}
\providecommand{\doarXiv}[1]{\href{https://arxiv.org/abs/#1}{\nolinkurl{https://arxiv.org/abs/#1}}}

\bibitem[{Ali {et~al.}(2015)Ali, Sittler, Chornay, Rowe, \&
  Puzzarini}]{Ali2015}
Ali, A., Sittler, E., Chornay, D., Rowe, B., \& Puzzarini, C. 2015, Planetary
  and Space Science, 109-110, 46 ,
  \dodoi{https://doi.org/10.1016/j.pss.2015.01.015}

\bibitem[{{Astropy Collaboration} {et~al.}(2013){Astropy Collaboration},
  {Robitaille}, {Tollerud}, {Greenfield}, {Droettboom}, {Bray}, {Aldcroft},
  {Davis}, {Ginsburg}, {Price-Whelan}, {Kerzendorf}, {Conley}, {Crighton},
  {Barbary}, {Muna}, {Ferguson}, {Grollier}, {Parikh}, {Nair}, {Unther},
  {Deil}, {Woillez}, {Conseil}, {Kramer}, {Turner}, {Singer}, {Fox}, {Weaver},
  {Zabalza}, {Edwards}, {Azalee Bostroem}, {Burke}, {Casey}, {Crawford},
  {Dencheva}, {Ely}, {Jenness}, {Labrie}, {Lim}, {Pierfederici}, {Pontzen},
  {Ptak}, {Refsdal}, {Servillat}, \& {Streicher}}]{2013A&A...558A..33A}
{Astropy Collaboration}, {Robitaille}, T.~P., {Tollerud}, E.~J., {et~al.} 2013,
  \aap, 558, A33, \dodoi{10.1051/0004-6361/201322068}

\bibitem[{{Astropy Collaboration} {et~al.}(2018){Astropy Collaboration},
  {Price-Whelan}, {Sip{\H{o}}cz}, {G{\"u}nther}, {Lim}, {Crawford}, {Conseil},
  {Shupe}, {Craig}, {Dencheva}, {Ginsburg}, {Vand erPlas}, {Bradley},
  {P{\'e}rez-Su{\'a}rez}, {de Val-Borro}, {Aldcroft}, {Cruz}, {Robitaille},
  {Tollerud}, {Ardelean}, {Babej}, {Bach}, {Bachetti}, {Bakanov}, {Bamford},
  {Barentsen}, {Barmby}, {Baumbach}, {Berry}, {Biscani}, {Boquien}, {Bostroem},
  {Bouma}, {Brammer}, {Bray}, {Breytenbach}, {Buddelmeijer}, {Burke},
  {Calderone}, {Cano Rodr{\'\i}guez}, {Cara}, {Cardoso}, {Cheedella}, {Copin},
  {Corrales}, {Crichton}, {D'Avella}, {Deil}, {Depagne}, {Dietrich}, {Donath},
  {Droettboom}, {Earl}, {Erben}, {Fabbro}, {Ferreira}, {Finethy}, {Fox},
  {Garrison}, {Gibbons}, {Goldstein}, {Gommers}, {Greco}, {Greenfield},
  {Groener}, {Grollier}, {Hagen}, {Hirst}, {Homeier}, {Horton}, {Hosseinzadeh},
  {Hu}, {Hunkeler}, {Ivezi{\'c}}, {Jain}, {Jenness}, {Kanarek}, {Kendrew},
  {Kern}, {Kerzendorf}, {Khvalko}, {King}, {Kirkby}, {Kulkarni}, {Kumar},
  {Lee}, {Lenz}, {Littlefair}, {Ma}, {Macleod}, {Mastropietro}, {McCully},
  {Montagnac}, {Morris}, {Mueller}, {Mumford}, {Muna}, {Murphy}, {Nelson},
  {Nguyen}, {Ninan}, {N{\"o}the}, {Ogaz}, {Oh}, {Parejko}, {Parley}, {Pascual},
  {Patil}, {Patil}, {Plunkett}, {Prochaska}, {Rastogi}, {Reddy Janga},
  {Sabater}, {Sakurikar}, {Seifert}, {Sherbert}, {Sherwood-Taylor}, {Shih},
  {Sick}, {Silbiger}, {Singanamalla}, {Singer}, {Sladen}, {Sooley},
  {Sornarajah}, {Streicher}, {Teuben}, {Thomas}, {Tremblay}, {Turner},
  {Terr{\'o}n}, {van Kerkwijk}, {de la Vega}, {Watkins}, {Weaver}, {Whitmore},
  {Woillez}, {Zabalza}, \& {Astropy Contributors}}]{2018AJ....156..123A}
{Astropy Collaboration}, {Price-Whelan}, A.~M., {Sip{\H{o}}cz}, B.~M., {et~al.}
  2018, \aj, 156, 123, \dodoi{10.3847/1538-3881/aabc4f}

\bibitem[{Berry {et~al.}(2019)Berry, Ugelow, Tolbert, \& Browne}]{Berry2019b}
Berry, J.~L., Ugelow, M.~S., Tolbert, M.~A., \& Browne, E.~C. 2019, \apj, 885,
  L6, \dodoi{10.3847/2041-8213/ab4b5b}

\bibitem[{Cable {et~al.}(2014)Cable, Hörst, He, Stockton, Mora, Tolbert,
  Smith, \& Willis}]{Cable2014}
Cable, M., Hörst, S., He, C., {et~al.} 2014, Earth and Planetary Science
  Letters, 403, 99 , \dodoi{https://doi.org/10.1016/j.epsl.2014.06.028}

\bibitem[{Caswell {et~al.}(2020)Caswell, Droettboom, Lee, Hunter, Firing,
  de~Andrade, Hoffmann, Stansby, Klymak, Varoquaux, Nielsen, Root, Elson, May,
  Dale, Lee, Seppänen, McDougall, Straw, Hobson, Gohlke, Yu, Ma, Vincent,
  Silvester, Moad, Kniazev, hannah, \& Ernest}]{matplotlib}
Caswell, T.~A., Droettboom, M., Lee, A., {et~al.} 2020, matplotlib/matplotlib:
  REL: v3.2.2, v3.2.2,  Zenodo, \dodoi{10.5281/zenodo.3898017}

\bibitem[{Christodoulides {et~al.}(1984)Christodoulides, McCorkle, \&
  Christophorou}]{christodoulides1984electron}
Christodoulides, A., McCorkle, D., \& Christophorou, L. 1984, Electron-Molecule
  Interactions and Their Applications, 2, 423

\bibitem[{Coates {et~al.}(2007)Coates, Crary, Lewis, Young, Waite~Jr., \&
  Sittler~Jr.}]{Coates2007}
Coates, A., Crary, F., Lewis, G., {et~al.} 2007, \grl, 34,
  \dodoi{10.1029/2007GL030978}

\bibitem[{Coates {et~al.}(2009)Coates, Wellbrock, Lewis, Jones, Young, Crary,
  \& Waite}]{Coates2009}
Coates, A., Wellbrock, A., Lewis, G., {et~al.} 2009, Planetary and Space
  Science, 57, 1866 , \dodoi{https://doi.org/10.1016/j.pss.2009.05.009}

\bibitem[{Coates {et~al.}(2010)Coates, Wellbrock, Lewis, Jones, Young, Crary,
  Waite, Johnson, Hill, \& Sittler~Jr.}]{Coates2010-RSC}
---. 2010, Faraday Discuss., 147, 293, \dodoi{10.1039/C004700G}

\bibitem[{Coates {et~al.}(2011)Coates, Wahlund, Ågren, Edberg, Cui, Wellbrock,
  \& Szego}]{Coates2011}
Coates, A.~J., Wahlund, J.-E., Ågren, K., {et~al.} 2011, Space Science
  Reviews, 162, 85, \dodoi{10.1007/s11214-011-9826-4}

\bibitem[{{Coustenis} {et~al.}(2003){Coustenis}, {Salama}, {Schulz}, {Ott},
  {Lellouch}, {Encrenaz}, {Gautier}, \& {Feuchtgruber}}]{Coustenis2003}
{Coustenis}, A., {Salama}, A., {Schulz}, B., {et~al.} 2003, \icarus, 161, 383,
  \dodoi{10.1016/S0019-1035(02)00028-3}

\bibitem[{{Coustenis} {et~al.}(2007){Coustenis}, {Achterberg}, {Conrath},
  {Jennings}, {Marten}, {Gautier}, {Nixon}, {Flasar}, {Teanby}, {B{\'e}zard},
  {Samuelson}, {Carlson}, {Lellouch}, {Bjoraker}, {Romani}, {Taylor}, {Irwin},
  {Fouchet}, {Hubert}, {Orton}, {Kunde}, {Vinatier}, {Mondellini}, {Abbas}, \&
  {Courtin}}]{Coustenis2007}
{Coustenis}, A., {Achterberg}, R.~K., {Conrath}, B.~J., {et~al.} 2007, Icarus,
  189, 35, \dodoi{10.1016/j.icarus.2006.12.022}

\bibitem[{Crary {et~al.}(2009)Crary, Magee, Mandt, Waite, Westlake, \&
  Young}]{Crary2009}
Crary, F., Magee, B., Mandt, K., {et~al.} 2009, Planetary and Space Science,
  57, 1847 , \dodoi{https://doi.org/10.1016/j.pss.2009.09.006}

\bibitem[{Cravens {et~al.}(2006)Cravens, Robertson, Waite~Jr., Yelle, Kasprzak,
  Keller, Ledvina, Niemann, Luhmann, McNutt, Ip, De~La~Haye, Mueller-Wodarg,
  Wahlund, Anicich, \& Vuitton}]{Cravens2006}
Cravens, T.~E., Robertson, I.~P., Waite~Jr., J.~H., {et~al.} 2006, \grl, 33,
  \dodoi{10.1029/2005GL025575}

\bibitem[{{Cui} {et~al.}(2009){Cui}, {Yelle}, {Vuitton}, {Waite}, {Kasprzak},
  {Gell}, {Niemann}, {M{\"u}ller-Wodarg}, {Borggren}, {Fletcher}, {Patrick},
  {Raaen}, \& {Magee}}]{Cui2009composition}
{Cui}, J., {Yelle}, R.~V., {Vuitton}, V., {et~al.} 2009, Icarus, 200, 581,
  \dodoi{10.1016/j.icarus.2008.12.005}

\bibitem[{Cui {et~al.}(2009)Cui, Galand, Yelle, Vuitton, Wahlund, Lavvas,
  Müller-Wodarg, Cravens, Kasprzak, \& Waite~Jr.}]{Cui2009}
Cui, J., Galand, M., Yelle, R.~V., {et~al.} 2009, Journal of Geophysical
  Research: Space Physics, 114, \dodoi{10.1029/2009JA014228}

\bibitem[{Dabestani \& Ivanov(1999)}]{Dabestani1999}
Dabestani, R., \& Ivanov, I.~N. 1999, Photochemistry and Photobiology, 70, 10,
  \dodoi{10.1111/j.1751-1097.1999.tb01945.x}

\bibitem[{Desai {et~al.}(2017)Desai, Coates, Wellbrock, Vuitton, Crary,
  Gonz{\'{a}}lez-Caniulef, Shebanits, Jones, Lewis, Waite, Cordiner, Taylor,
  Kataria, Wahlund, Edberg, \& Sittler}]{Desai_2017}
Desai, R., Coates, A., Wellbrock, A., {et~al.} 2017, \apj, 844, L18,
  \dodoi{10.3847/2041-8213/aa7851}

\bibitem[{Dubois {et~al.}(2020)Dubois, Carrasco, Jovanovic, Vettier, Gautier,
  \& Westlake}]{Dubois2020}
Dubois, D., Carrasco, N., Jovanovic, L., {et~al.} 2020, Icarus, 338, 113437,
  \dodoi{https://doi.org/10.1016/j.icarus.2019.113437}

\bibitem[{{Fleury} {et~al.}(2014){Fleury}, {Carrasco}, {Gautier}, {Mahjoub},
  {He}, {Szopa}, {Hadamcik}, {Buch}, \& {Cernogora}}]{Fleury2014}
{Fleury}, B., {Carrasco}, N., {Gautier}, T., {et~al.} 2014, Icarus, 238, 221,
  \dodoi{10.1016/j.icarus.2014.05.027}

\bibitem[{{Gautier} {et~al.}(2017){Gautier}, {Sebree}, {Li}, {Pinnick},
  {Grubisic}, {Loeffler}, {Getty}, {Trainer}, \& {Brinckerhoff}}]{Gautier2017}
{Gautier}, T., {Sebree}, J.~A., {Li}, X., {et~al.} 2017, \planss, 140, 27,
  \dodoi{10.1016/j.pss.2017.03.012}

\bibitem[{{Gurnett} {et~al.}(2004){Gurnett}, {Kurth}, {Kirchner},
  {Hospodarsky}, {Averkamp}, {Zarka}, {Lecacheux}, {Manning}, {Roux}, {Canu},
  {Cornilleau-Wehrlin}, {Galopeau}, {Meyer}, {Bostr{\"o}m}, {Gustafsson},
  {Wahlund}, {{\r{A}}hlen}, {Rucker}, {Ladreiter}, {Macher}, {Woolliscroft},
  {Alleyne}, {Kaiser}, {Desch}, {Farrell}, {Harvey}, {Louarn}, {Kellogg},
  {Goetz}, \& {Pedersen}}]{Gurnett2004}
{Gurnett}, D.~A., {Kurth}, W.~S., {Kirchner}, D.~L., {et~al.} 2004, \ssr, 114,
  395, \dodoi{10.1007/s11214-004-1434-0}

\bibitem[{{Hartle} {et~al.}(2006){Hartle}, {Sittler}, {Neubauer}, {Johnson},
  {Smith}, {Crary}, {McComas}, {Young}, {Coates}, {Simpson}, {Bolton},
  {Reisenfeld}, {Szego}, {Berthelier}, {Rymer}, {Vilppola}, {Steinberg}, \&
  {Andre}}]{Hartle2006}
{Hartle}, R.~E., {Sittler}, E.~C., {Neubauer}, F.~M., {et~al.} 2006, \grl, 33,
  L08201, \dodoi{10.1029/2005GL024817}

\bibitem[{{H{\"o}rst} {et~al.}(2012){H{\"o}rst}, {Yelle}, {Buch}, {Carrasco},
  {Cernogora}, {Dutuit}, {Quirico}, {Sciamma-O'Brien}, {Smith}, {Somogyi},
  {Szopa}, {Thissen}, \& {Vuitton}}]{Horst2012}
{H{\"o}rst}, S.~M., {Yelle}, R.~V., {Buch}, A., {et~al.} 2012, Astrobiology,
  12, 809, \dodoi{10.1089/ast.2011.0623}

\bibitem[{{Lavvas} {et~al.}(2011){Lavvas}, {Sander}, {Kraft}, \&
  {Imanaka}}]{Lavvas2011}
{Lavvas}, P., {Sander}, M., {Kraft}, M., \& {Imanaka}, H. 2011, \apj, 728, 80,
  \dodoi{10.1088/0004-637X/728/2/80}

\bibitem[{{Lavvas} {et~al.}(2013){Lavvas}, {Yelle}, {Koskinen}, {Bazin},
  {Vuitton}, {Vigren}, {Galand }, {Wellbrock}, {Coates}, {Wahlund}, {Crary}, \&
  {Snowden}}]{Lavvas2013}
{Lavvas}, P., {Yelle}, R.~V., {Koskinen}, T., {et~al.} 2013, Proceedings of the
  National Academy of Science, 110, 2729, \dodoi{10.1073/pnas.1217059110}

\bibitem[{{Lavvas} {et~al.}(2008{\natexlab{a}}){Lavvas}, {Coustenis}, \&
  {Vardavas}}]{Lavvas2008a}
{Lavvas}, P.~P., {Coustenis}, A., \& {Vardavas}, I.~M. 2008{\natexlab{a}},
  \planss, 56, 27, \dodoi{10.1016/j.pss.2007.05.026}

\bibitem[{{Lavvas} {et~al.}(2008{\natexlab{b}}){Lavvas}, {Coustenis}, \&
  {Vardavas}}]{Lavvas2008b}
---. 2008{\natexlab{b}}, \planss, 56, 67, \dodoi{10.1016/j.pss.2007.05.027}

\bibitem[{{Le Page} {et~al.}(1999){Le Page}, {Keheyan}, {Snow}, \&
  {Bierbaum}}]{LePage1999a}
{Le Page}, V., {Keheyan}, Y., {Snow}, T.~P., \& {Bierbaum}, V.~M. 1999,
  International Journal of Mass Spectrometry, 185, 949,
  \dodoi{10.1016/S1387-3806(98)14217-3}

\bibitem[{{Le Page} {et~al.}(1999b){Le Page}, {Keheyan}, {Snow}, \&
  {Bierbaum}}]{LePage1999b}
---. 1999b, International Journal of Mass Spectrometry, 185, 949,
  \dodoi{10.1016/S1387-3806(98)14217-3}

\bibitem[{Lebonnois {et~al.}(2002)Lebonnois, Bakes, \& McKay}]{Lebonnis2002}
Lebonnois, S., Bakes, E., \& McKay, C.~P. 2002, Icarus, 159, 505 ,
  \dodoi{https://doi.org/10.1006/icar.2002.6943}

\bibitem[{{Lindal} {et~al.}(1983){Lindal}, {Wood}, {Hotz}, {Sweetnam},
  {Eshleman}, \& {Tyler}}]{Lindal1981}
{Lindal}, G.~F., {Wood}, G.~E., {Hotz}, H.~B., {et~al.} 1983, \icarus, 53, 348,
  \dodoi{10.1016/0019-1035(83)90155-0}

\bibitem[{{L{\'o}pez-Puertas} {et~al.}(2013){L{\'o}pez-Puertas}, {Dinelli},
  {Adriani}, {Funke}, {Garc{\'\i}a-Comas}, {Moriconi}, {D'Aversa}, {Boersma},
  \& {Allamandola}}]{Lopez-Puertas2013}
{L{\'o}pez-Puertas}, M., {Dinelli}, B.~M., {Adriani}, A., {et~al.} 2013, \apj,
  770, 132, \dodoi{10.1088/0004-637X/770/2/132}

\bibitem[{{Mahjoub} {et~al.}(2016){Mahjoub}, {Schwell}, {Carrasco}, {Benilan},
  {Cernogora}, {Szopa}, \& {Gazeau}}]{Mahjoub2016}
{Mahjoub}, A., {Schwell}, M., {Carrasco}, N., {et~al.} 2016, \planss, 131, 1,
  \dodoi{10.1016/j.pss.2016.05.003}

\bibitem[{Mandt {et~al.}(2012)Mandt, Gell, Perry, Hunter Waite~Jr., Crary,
  Young, Magee, Westlake, Cravens, Kasprzak, Miller, Wahlund, Ågren, Edberg,
  Heays, Lewis, Gibson, de~la Haye, \& Liang}]{Mandt2012}
Mandt, K.~E., Gell, D.~A., Perry, M., {et~al.} 2012, Journal of Geophysical
  Research: Planets, 117, \dodoi{10.1029/2012JE004139}

\bibitem[{{Mihailescu} {et~al.}(2020){Mihailescu}, {Desai}, {Shebanits},
  {Haythornthwaite}, {Wellbrock}, {Coates}, {Eastwood}, \&
  {Waite}}]{Mihailescu2020}
{Mihailescu}, T., {Desai}, R.~T., {Shebanits}, O., {et~al.} 2020, The Planetary
  Science Journal, 1, 50, \dodoi{10.3847/PSJ/abb1ba}

\bibitem[{{Porco} {et~al.}(2005){Porco}, {Baker}, {Barbara}, {Beurle},
  {Brahic}, {Burns}, {Charnoz}, {Cooper}, {Dawson}, {Del Genio}, {Denk},
  {Dones}, {Dyudina}, {Evans}, {Fussner}, {Giese}, {Grazier}, {Helfenstein},
  {Ingersoll}, {Jacobson}, {Johnson}, {McEwen}, {Murray}, {Neukum}, {Owen},
  {Perry}, {Roatsch}, {Spitale}, {Squyres}, {Thomas}, {Tiscareno}, {Turtle},
  {Vasavada}, {Veverka}, {Wagner}, \& {West}}]{Porco2005}
{Porco}, C.~C., {Baker}, E., {Barbara}, J., {et~al.} 2005, \nat, 434, 159,
  \dodoi{10.1038/nature03436}

\bibitem[{{Sagan} {et~al.}(1993){Sagan}, {Khare}, {Thompson}, {McDonald},
  {Wing}, {Bada}, {Vo-Dinh}, \& {Arakawa}}]{Sagan1993}
{Sagan}, C., {Khare}, B.~N., {Thompson}, W.~R., {et~al.} 1993, \apj, 414, 399,
  \dodoi{10.1086/173086}

\bibitem[{{Sciamma-O'Brien} {et~al.}(2017){Sciamma-O'Brien}, {Upton}, \&
  {Salama}}]{SciammaOBrien2017}
{Sciamma-O'Brien}, E., {Upton}, K.~T., \& {Salama}, F. 2017, \icarus, 289, 214,
  \dodoi{10.1016/j.icarus.2017.02.004}

\bibitem[{Sittler {et~al.}(2020)Sittler, Cooper, Sturner, \& Ali}]{Sittler2020}
Sittler, E.~C., Cooper, J.~F., Sturner, S.~J., \& Ali, A. 2020, Icarus, 344,
  113246, \dodoi{https://doi.org/10.1016/j.icarus.2019.03.023}

\bibitem[{{Skouteris} {et~al.}(2015){Skouteris}, {Balucani}, {Faginas-Lago},
  {Falcinelli}, \& {Rosi}}]{Skouteris2015}
{Skouteris}, D., {Balucani}, N., {Faginas-Lago}, N., {Falcinelli}, S., \&
  {Rosi}, M. 2015, \aap, 584, A76, \dodoi{10.1051/0004-6361/201526978}

\bibitem[{{Trainer} {et~al.}(2013){Trainer}, {Sebree}, {Yoon}, \&
  {Tolbert}}]{Trainer2013}
{Trainer}, M.~G., {Sebree}, J.~A., {Yoon}, Y.~H., \& {Tolbert}, M.~A. 2013,
  \apjl, 766, L4, \dodoi{10.1088/2041-8205/766/1/L4}

\bibitem[{{Tyler} {et~al.}(1981){Tyler}, {Eshleman}, {Anderson}, {Levy},
  {Lindal}, {Wood}, \& {Croft}}]{Tyler1981}
{Tyler}, G.~L., {Eshleman}, V.~R., {Anderson}, J.~D., {et~al.} 1981, Science,
  212, 201, \dodoi{10.1126/science.212.4491.201}

\bibitem[{{van der Walt} {et~al.}(2011){van der Walt}, {Colbert}, \&
  {Varoquaux}}]{2011CSE....13b..22V}
{van der Walt}, S., {Colbert}, S.~C., \& {Varoquaux}, G. 2011, Computing in
  Science and Engineering, 13, 22, \dodoi{10.1109/MCSE.2011.37}

\bibitem[{{Vilppola} {et~al.}(2001){Vilppola}, {Tanskanen}, {Barraclough}, \&
  {McComas}}]{Vilppola2001}
{Vilppola}, J.~H., {Tanskanen}, P.~J., {Barraclough}, B.~L., \& {McComas},
  D.~J. 2001, Review of Scientific Instruments, 72, 3662,
  \dodoi{10.1063/1.1392337}

\bibitem[{{Virtanen} {et~al.}(2020){Virtanen}, {Gommers}, {Oliphant},
  {Haberland}, {Reddy}, {Cournapeau}, {Burovski}, {Peterson}, {Weckesser},
  {Bright}, {van der Walt}, {Brett}, {Wilson}, {Jarrod Millman}, {Mayorov},
  {Nelson}, {Jones}, {Kern}, {Larson}, {Carey}, {Polat}, {Feng}, {Moore}, {Vand
  erPlas}, {Laxalde}, {Perktold}, {Cimrman}, {Henriksen}, {Quintero}, {Harris},
  {Archibald}, {Ribeiro}, {Pedregosa}, {van Mulbregt}, \&
  {Contributors}}]{2020SciPy-NMeth}
{Virtanen}, P., {Gommers}, R., {Oliphant}, T.~E., {et~al.} 2020, Nature
  Methods, 17, 261, \dodoi{https://doi.org/10.1038/s41592-019-0686-2}

\bibitem[{Vuitton {et~al.}(2019)Vuitton, Yelle, Klippenstein, Hörst, \&
  Lavvas}]{Vuitton2019}
Vuitton, V., Yelle, R., Klippenstein, S., Hörst, S., \& Lavvas, P. 2019,
  Icarus, 324, 120 , \dodoi{https://doi.org/10.1016/j.icarus.2018.06.013}

\bibitem[{Vuitton {et~al.}(2007)Vuitton, Yelle, \& McEwan}]{Vuitton2007}
Vuitton, V., Yelle, R., \& McEwan, M. 2007, Icarus, 191, 722 ,
  \dodoi{https://doi.org/10.1016/j.icarus.2007.06.023}

\bibitem[{{Vuitton} {et~al.}(2006){Vuitton}, {Yelle}, \&
  {Anicich}}]{Vuitton2006}
{Vuitton}, V., {Yelle}, R.~V., \& {Anicich}, V.~G. 2006, \apjl, 647, L175,
  \dodoi{10.1086/507467}

\bibitem[{Waite {et~al.}(2007)Waite, Young, Cravens, Coates, Crary, Magee, \&
  Westlake}]{Waite2007}
Waite, J.~H., Young, D.~T., Cravens, T.~E., {et~al.} 2007, Science, 316, 870,
  \dodoi{10.1126/science.1139727}

\bibitem[{Waite {et~al.}(2005)Waite, Niemann, Yelle, Kasprzak, Cravens,
  Luhmann, McNutt, Ip, Gell, De~La~Haye, M{\"u}ller-Wordag, Magee, Borggren,
  Ledvina, Fletcher, Walter, Miller, Scherer, Thorpe, Xu, Block, \&
  Arnett}]{Waite2005}
Waite, J.~H., Niemann, H., Yelle, R.~V., {et~al.} 2005, Science, 308, 982,
  \dodoi{10.1126/science.1110652}

\bibitem[{Wellbrock {et~al.}(2013)Wellbrock, Coates, Jones, Lewis, \&
  Waite}]{Wellbrock2013}
Wellbrock, A., Coates, A., Jones, G., Lewis, G., \& Waite, J. 2013, \grl, 40,
  4481, \dodoi{10.1002/grl.50751}

\bibitem[{Wellbrock {et~al.}(2012)Wellbrock, {Coates}, {Jones}, {Arridge},
  {Lewis}, {Sittler}, \& {Young}}]{Wellbrock2012}
Wellbrock, A., {Coates}, A.~J., {Jones}, G.~H., {et~al.} 2012, in AGU Fall
  Meeting Abstracts, Vol. 2012, P21F--1898

\bibitem[{Wellbrock {et~al.}(2019)Wellbrock, Coates, Jones, Vuitton, Lavvas,
  Desai, \& Waite}]{Wellbrock2019}
Wellbrock, A., Coates, A.~J., Jones, G.~H., {et~al.} 2019, \mnras, 490, 2254,
  \dodoi{10.1093/mnras/stz2655}

\bibitem[{Wellbrock \& et~al.(2021, in prep)}]{Wellbrock2020}
Wellbrock, A., \& et~al. 2021, in prep

\bibitem[{Wellbrock {et~al.}(2009)Wellbrock, {Coates}, {Jones}, {Arridge},
  {Lewis}, {Magee}, {Waite}, {Sittler}, {Crary}, \& {Young}}]{Wellbrock2009}
Wellbrock, A., {Coates}, A.~J., {Jones}, G.~H., {et~al.} 2009, in AGU Fall
  Meeting Abstracts, Vol. 2009, P51G--1198

\bibitem[{Westlake {et~al.}(2011)Westlake, Bell, Waite~Jr., Johnson, Luhmann,
  Mandt, Magee, \& Rymer}]{Westlake2011}
Westlake, J.~H., Bell, J.~M., Waite~Jr., J.~H., {et~al.} 2011, Journal of
  Geophysical Research: Space Physics, 116, \dodoi{10.1029/2010JA016251}

\bibitem[{Westlake {et~al.}(2014)Westlake, Waite~Jr., Carrasco, Richard, \&
  Cravens}]{Westlake2014}
Westlake, J.~H., Waite~Jr., J.~H., Carrasco, N., Richard, M., \& Cravens, T.
  2014, {Journal of Geophysical Research: Space Physics}, 119, 5951,
  \dodoi{10.1002/2014JA020208}

\bibitem[{Wilson \& Atreya(2003)}]{Wilson2003}
Wilson, E., \& Atreya, S. 2003, Planetary and Space Science, 51, 1017 ,
  \dodoi{https://doi.org/10.1016/j.pss.2003.06.003}

\bibitem[{{Yoon} {et~al.}(2014){Yoon}, {H{\"o}rst}, {Hicks}, {Li}, {de Gouw},
  \& {Tolbert}}]{Yoon2014}
{Yoon}, Y.~H., {H{\"o}rst}, S.~M., {Hicks}, R.~K., {et~al.} 2014, \icarus, 233,
  233, \dodoi{10.1016/j.icarus.2014.02.006}

\bibitem[{Young {et~al.}(2004)Young, Berthelier, Blanc, Burch, Coates,
  Goldstein, Grande, Hill, Johnson, Kelha, Mccomas, Sittler, Svenes, Szeg{\"o},
  Tanskanen, Ahola, Anderson, Bakshi, Baragiola, Barraclough, Black, Bolton,
  Booker, Bowman, Casey, Crary, Delapp, Dirks, Eaker, Funsten, Furman, Gosling,
  Hannula, Holmlund, Huomo, Illiano, Jensen, Johnson, Linder, Luntama, Maurice,
  Mccabe, Mursula, Narheim, Nordholt, Preece, Rudzki, Ruitberg, Smith, Szalai,
  Thomsen, Viherkanto, Vilppola, Vollmer, Wahl, W{\"u}est, Ylikorpi, \&
  Zinsmeyer}]{Young2004}
Young, D., Berthelier, J., Blanc, M., {et~al.} 2004, Space Science Reviews,
  114, 1, \dodoi{10.1007/s11214-004-1406-4}

\bibitem[{{Zhao} {et~al.}(2018){Zhao}, {Kaiser}, {Xu}, {Ablikim}, {Ahmed},
  {Evseev}, {Bashkirov}, {Azyazov}, \& {Mebel}}]{Zhao2018}
{Zhao}, L., {Kaiser}, R.~I., {Xu}, B., {et~al.} 2018, Nature Astronomy, 2, 973,
  \dodoi{10.1038/s41550-018-0585-y}

\bibitem[{Ågren {et~al.}(2009)Ågren, Wahlund, Garnier, Modolo, Cui, Galand,
  \& Müller-Wodarg}]{Agren2009}
Ågren, K., Wahlund, J.~E., Garnier, P., {et~al.} 2009, Planetary and Space
  Science, 57, 1821, \dodoi{10.1016/j.pss.2009.04.012}

\end{thebibliography}
\bibliographystyle{aasjournal}

\listofchanges

\end{document}